\documentclass[12pt,a4paper,oneside]{article}
\usepackage{a4wide}
\usepackage[T1]{fontenc}
\usepackage[utf8]{inputenc}
\usepackage{amsmath}
\usepackage{times}
\usepackage{ulem}
\usepackage{graphicx}
\usepackage[final]{pdfpages}
\usepackage{longtable}
\usepackage{float}
\usepackage{amssymb}
\usepackage{url}
\usepackage{xcolor}
\usepackage{mathtools}
\usepackage{tcolorbox}
\definecolor{pigpink}{HTML}{FDD7E4}
\definecolor{lcyan}{HTML}{E0FFFF}
\definecolor{mint}{HTML}{98FF98}

\usepackage{marginnote}
\usepackage{soul}

\usepackage{amsthm}
\newtheorem{theorem}{Theorem}[section]

\newtheorem{remark}[theorem]{Remark}

\usepackage{indentfirst}
\title{\vspace{-1cm}Phase-locked states in oscillating neural networks and their role in neural communication}

\author{Alberto P\'erez-Cervera$^{1}$, Tere M. Seara$^{1}$ and Gemma Huguet$^{1}$\\
	\parbox{12.5cm}{
		\small
		\begin{itemize}
			\item[$^1$]
			Departament de Matem\`atiques, Universitat Polit\`ecnica de
			Catalunya, Avda. Diagonal 647, 08028 Barcelona. BGSMATH \\
		\end{itemize}
	}}

\begin{document}

		%----------------------------------------------------------------------------
		%\oddsidemargin -20pt
		% \begin{titlepage}
		%
		% \vspace*{3cm}
		% \centerline{Defensa de Pla de Recerca}
		% \bigskip
		% \centerline{\Large \bf Computing PRCs across the}
		% \centerline{\Large \bf Parameterization Method}
		% \bigskip
		% \centerline{Alberto PÃ©rez Cervera}
		% \bigskip
		%
		% \bigskip
		% \end{titlepage}
		%
		% \newpage
		% \thispagestyle{empty} % para que no se numere esta pagina
		% $\ $
		% \newpage{\pagestyle{empty}\cleardoublepage}
		% \thispagestyle{empty} % para que no se numere esta pagina

		%=====================================================================================
		
		\date{}
		
		\maketitle
		
		\noindent \textbf{Corresponding author:} Alberto P\'erez-Cervera,
		\texttt{alberto.perez@upc.edu} \\
		
		
		%37D10 invariant manifold theory
		%92B25 Biological rhythms and synchronization
		%65P30 Bifurcation problems
		%37N25 Dynamical systems in biology 
		\noindent

		\section*{Abstract}
		
The theory of communication through coherence (CTC) proposes that brain oscillations reflect changes in the excitability of neurons, and therefore the successful communication between two oscillating neural populations depends not only on the strength of the signal emitted but also on the relative phases between them. More precisely, effective communication occurs when the emitting and receiving populations are properly phase locked so the inputs sent by the emitting population arrive at the phases of maximal excitability of the receiving population. To study this setting, we consider a population rate model consisting of excitatory and inhibitory cells modelling the receiving population, and we perturb it with a time-dependent periodic function modelling the input from the emitting population. We consider the stroboscopic map for this system and compute numerically the fixed and periodic points of this map and their bifurcations as the amplitude and the frequency of the perturbation are varied. From the bifurcation diagram, we identify the phase-locked states as well as different regions of bistability. We explore carefully the dynamics of particular phase-locking regimes emphasizing its implications for the CTC theory. In particular, we study how the input gain  depends on the timing between the input and the inhibitory action of the receiving population. Our results show that naturally an optimal phase locking for CTC emerges, and provide a mechanism by which the receiving population can implement selective communication. Moreover, the presence of bistable regions, suggests a mechanism by which different communication regimes between brain areas can be established without changing the structure of the network. \\

\textbf{List of abbreviations}\\

\begin{tabular}{l l}
	CTC & Communication Through Coherence\\
	E-I & Excitatory-Inhibitory
\end{tabular}
		
\newpage		

\section{Introduction}

Neural oscillations are ubiquitous in the brain. Since they were first observed in 1929 by Hans
Berger~\cite{berger1929}, they have been profusely studied to unveil their link with brain function. Nowadays, they 
are classified in the following bands: delta (1-4 Hz), theta (4-8 Hz), alpha (8-13 Hz), beta (13-30 Hz) and gamma (30-70 Hz). Although some of these frequency bands have been associated to specific tasks or behaviours, their functional role is still not completely understood \cite{buzsaki2006rhythms}. 

\begin{figure}[b]
	\begin{minipage}[c]{0.5\textwidth}
		\includegraphics[width=80mm]{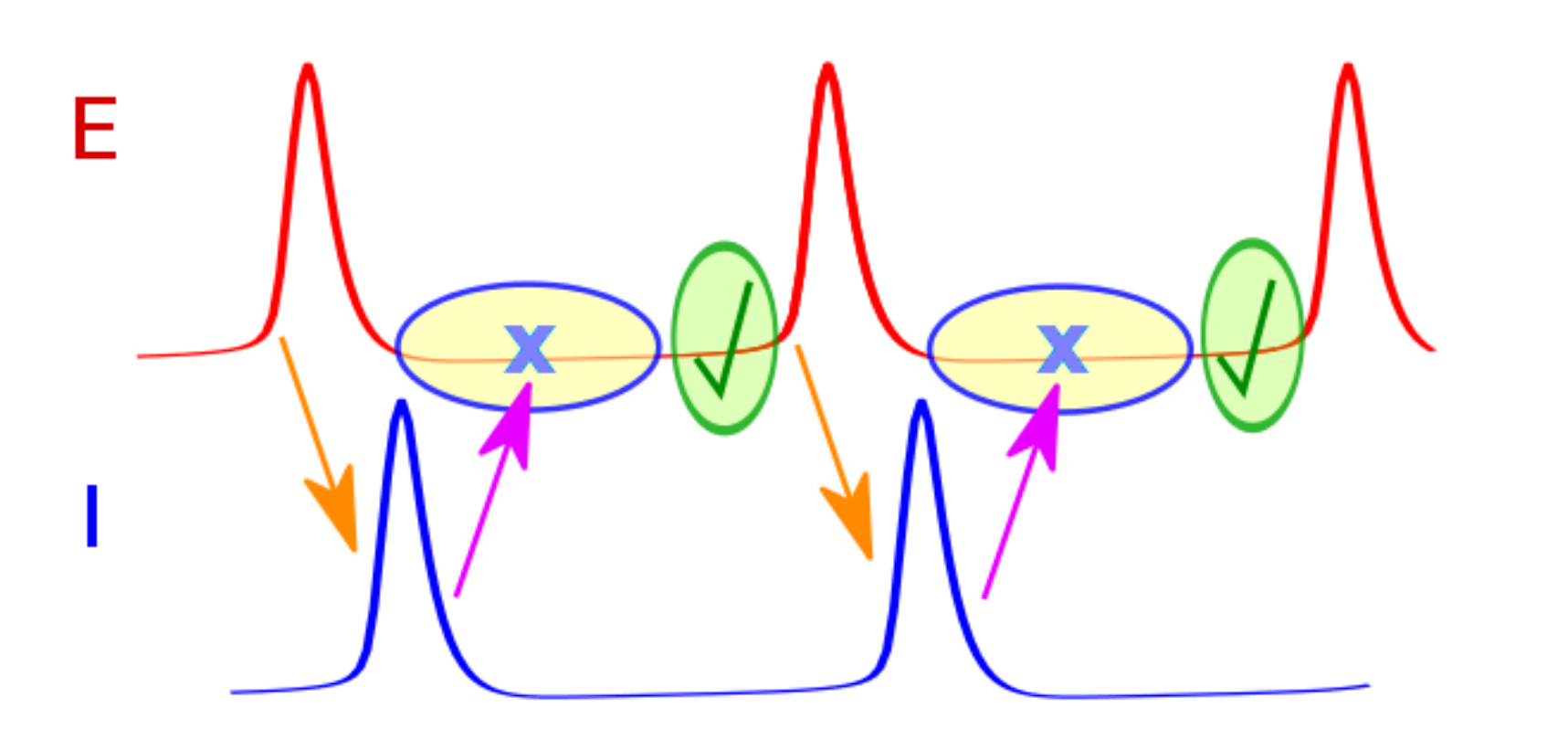}
	\end{minipage}\hfill
	\begin{minipage}[c]{0.45\textwidth}
		\caption{The picture illustrates different excitability properties along a cycle generated by the interaction between excitation (red) and inhibition (blue). Once the inhibition decays, the network is sensitive to external inputs.} \label{fig:ctcFramework}
	\end{minipage}
\end{figure}

Fast brain oscillations in the gamma frequency band have been hypothesized to occur in local neural networks composed by excitatory pyramidal neurons and inhibitory interneurons (E-I networks) \cite{buzsaki2012mechanisms, bartos2007synaptic}. There is an increasing number of studies which link oscillations in the gamma band frequencies with cognitive processes and communication between brain areas \cite{buzsaki2004neuronal, fries2007gamma,tiesinga2009cortical}. In this context, the Communication Through Coherence (CTC) Theory \cite{fries2005} conjectures that oscillations can account for a flexible mechanism of communication between neural populations. More precisely, oscillations generated across the interaction of excitatory and inhibitory cells cause that the excitability of the excitatory population is not the same for all the phases of the 
cycle due to the inhibitory action \cite{kopell2000gamma, tiesinga2001computational}. Indeed, when the excitatory population receives an external input at the phase in which the inhibition is not present, the excitatory cells can 
respond effectively, thus promoting communication, while if the inhibition is present, the input might be ignored, thus preventing communication (see Fig.~\ref{fig:ctcFramework}). Therefore, according to the CTC theory, two neuronal populations with underlying oscillatory activity communicate much effectively when they are coherent, that is, they are properly phase locked so that the output sent by the pre-synaptic (emitting) population reaches the post-synaptic (receiving) population in its peaks of excitability.

Different predictions following the CTC theory have been experimentally tested \cite{fries2015rhythms}. On one hand, different studies link the phase of the inhibitory receiving population with the modulation of the input gain \cite{cardin2009driving}. On the other hand, different studies support that selective communication, that is, the ability of the post-synaptic group to respond to a given input and ignore the others, is implemented through selective coherence \cite{schoffelen2011selective,fries2001}.

Furthermore, besides the experimental studies, the CTC framework has also been studied by means of computational models, which focus on the link between gamma oscillations and stimulus selection \cite{tiesinga2010, borgers2008gamma, GielenKZ10}. Conclusions agree with the CTC hypothesis that the phase relationship which is established between a rhythmic input and the post-synaptic group turns to be optimal for the CTC scheme \cite{cannon2014neurosystems}. Most of these computational studies are based on E-I networks of spiking neurons. Nevertheless, mean field approaches are also useful to complement these results and to gain insight into the mechanisms underlying the CTC hypotheses. Indeed, they allow for a more manageable analytical treatment, while keeping the essential processes involved \cite{breakspear2017dynamic}.

%In this work we propose a mean field approach to the CTC theory.
In this work we propose a theoretical approach to the CTC theory by means of a phenomenological description of the population activity in terms of the mean firing rate. More precisely, in order to deepen in the mechanism underlying phase-locking in a neuronal cycle having different excitability phases, 
we will consider the effect of an external periodic input onto a network model consisting of a single population of excitatory neurons and a single population of inhibitory neurons (E-I network). For such setting we consider the simplest canonical model describing the mean firing rates of an E-I network: 
the Wilson-Cowan equations~\cite{wilson1972}. The parameters of this model will be chosen so that the system shows oscillations~\cite{borisyuk1992}.
In particular, we focus on oscillations of the Wilson-Cowan model arising from a Hopf bifurcation -we also provide a preliminary exploration of the case close to a Saddle-Node on an Invariant Circle (SNIC) bifurcation. The goal is to study the different phase-locking patterns that emerge between the oscillatory E-I network and the external periodic input for different input parameters. We remark that our setting explores the dynamics emerging from unidirectional communication. Indeed, some studies have conjectured that a given brain area has neurons receiving inputs and different neurons sending outputs \cite{Markov14}. Nevertheless, we point out that schemes based on bidirectional communication have also been proposed to play a role 
in the context of CTC theory \cite{witt2013controlling, battaglia2012dynamic}. 

To determine and compute the phase-locked states, we consider the stroboscopic map for this system and compute numerically its fixed and periodic points and their bifurcations, as the amplitude and the frequency of the perturbation are varied. 
The techniques that we use to do the bifurcation analysis have no restriction neither on the amplitude nor on the frequency of the perturbation, or how close the limit cycle of the Wilson-Cowan equations is from a bifurcation. From the bifurcation diagram, we can 
identify the phase-locked states as well as different regions of bistability between different invariant objects. We explore carefully the dynamics on these invariant objects and we discuss the implications of these results for the CTC theory,
paying attention to the phase-locking and amplitude of the response of the oscillatory neuronal population to the external input.
Notably, our results provide a mechanism by which the receiving population can implement selective communication, as well as a mechanism by which different communication regimes between areas can be established (communication can be turned on and off) 
without changing the connectivity of the network. 
 
The structure of the paper is as follows. In Section \ref{sec:ctcApproach}, we introduce the mathematical model by which we explore the theoretical basis of CTC. Section \ref{sec:dyn_anal} contains the mathematical analysis of the model. More precisely, in Section \ref{sec:strbrMap}, we introduce the stroboscopic map and compute the bifurcation diagram of its fixed points as the frequency and the amplitude of the perturbation are varied. In Section \ref{sec:bifAnalisys} we provide a complete dynamical analysis of the different phase-locking regions including the bistability regions identified in the bifurcation diagram. In Section~\ref{sec:section4} we discuss the implications for CTC theory of the different dynamical scenarios found in Section \ref{sec:bifAnalisys} and we finish with a discussion in Section~\ref{sec:discussion}.
The Appendix~\ref{sec:bifAnalysis} contains details of the numerical algorithms used to compute the bifurcation diagram, Appendix~\ref{ap:bd_snic} contains a preliminary exploration of the Wilson-Cowan equations in the oscillatory regime close to a SNIC bifurcation and Appendix~\ref{ap:nonSin} includes an exploratory result for non-sinusoidal type of inputs.

\section{Mathematical setting for CTC}\label{sec:ctcApproach}

In this Section we present the theoretical setting based on a canonical population firing rate model that implements mathematically the CTC framework. We consider the classical  Wilson-Cowan model, which describes the behaviour of a coupled network of excitatory and inhibitory neurons \cite{wilson1972}, and we perturb it with a time-periodic function $p(t)$ which models the input from an external oscillating source. The perturbed Wilson-Cowan equations have the form
\begin{equation}\label{eq:WCsys}
\begin{split}
\dot{r_e} &= -r_e + S_e(c_1 r_e - c_2 r_i + P + Ap(t)),\\
\dot{r_i} &= -r_i + S_i(c_3 r_e - c_4 r_i + Q),
\end{split}
\end{equation}
where the variables $r_e$ and $r_i$ are the firing rate activity of the excitatory and inhibitory populations, respectively, and
\begin{equation}
S_{k}(x) = \frac{1}{1+ e^{-a_k(x-\theta_k)}}, \quad \textrm{for } k=e,i,
\end{equation}
is the input-output function.

\begin{figure}[t]
	\begin{minipage}[c]{0.69\textwidth}
		{\includegraphics[width=110mm]{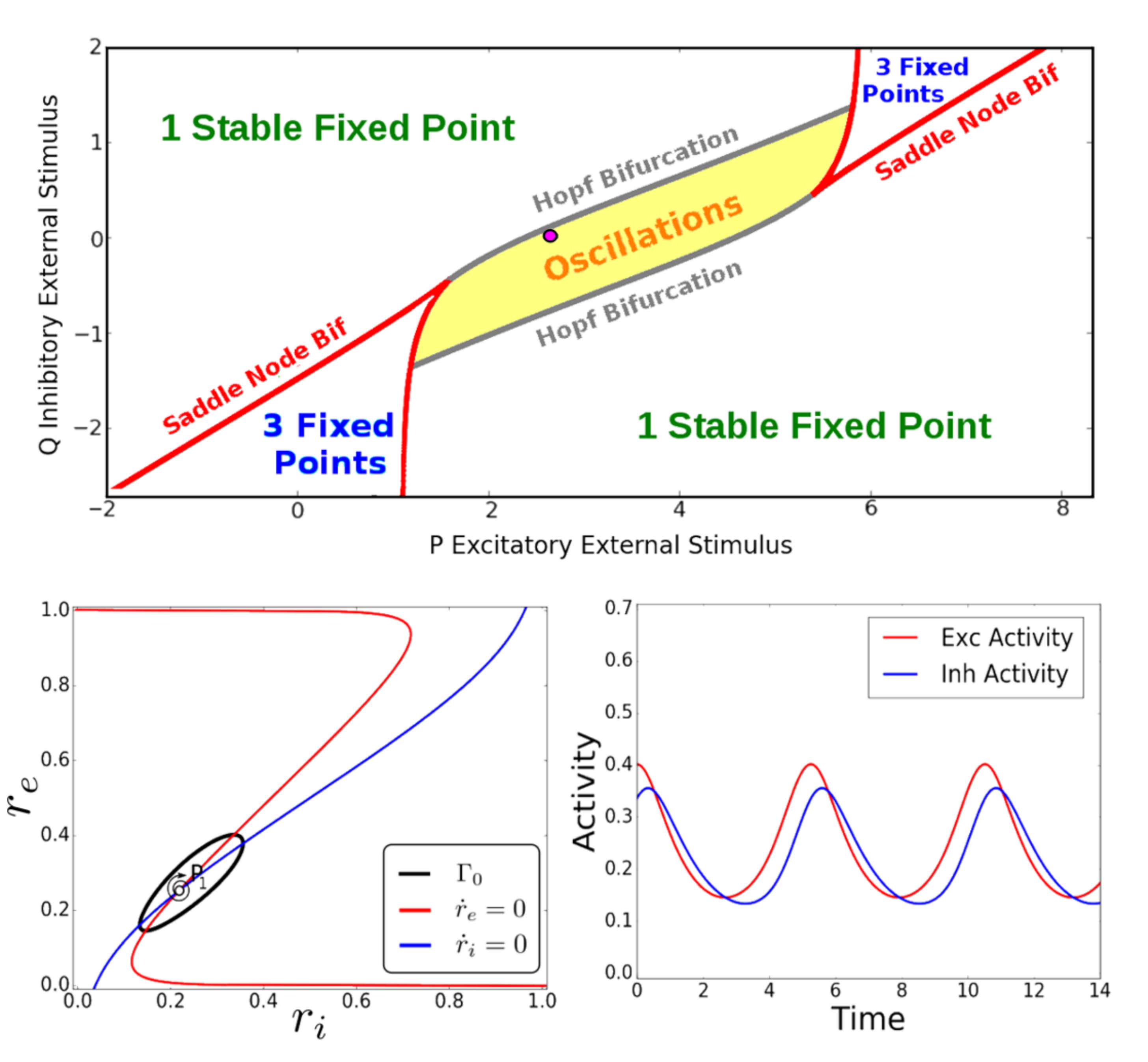}}
	\end{minipage}\hfill
	\begin{minipage}[c]{0.31\textwidth}
		\caption{For the unperturbed ($A=0$) Wilson-Cowan equations \eqref{eq:WCsys} having the set of parameters $\mathcal{P}$ given in \eqref{eq:parametersChoice} we show: Bifurcation diagram as a function of the external stimuli $P$ and $Q$ (Top panel). The pink dot indicates the pair of ($P, Q$) = (2.5, 0) values chosen so that \eqref{eq:WCsys} shows oscillations. The bottom-left panel shows the nullclines and the phase space for the choice ($P, Q$) = (2.5, 0). The phase space shows a limit cycle $\Gamma_0$ and an unstable focus $P_1$. The bottom-right panel shows the dynamics over the limit cycle $\Gamma_0$. Notice how oscillations arise from the interaction between excitatory and inhibitory activity.} \label{fig:2d_bifDiag}
	\end{minipage}
\end{figure}

As we will use the Wilson-Cowan equations to model the receiving population, we need to choose parameters in \eqref{eq:WCsys} such that for $A=0$ they show a limit cycle. The conditions for the Wilson-Cowan equations to display oscillations have been studied in classical papers \cite{wilson1972, borisyuk1992}. Typically, the external currents $P$ and $Q$ are set as the bifurcation parameters. The reason is because they translate the nullclines of system \eqref{eq:WCsys} and thus determine the position and number of the critical points. For this problem we will use the following set of parameters 
\begin{equation}\label{eq:parametersChoice}
\begin{split}
\mathcal{P} = \{ c_{1}=13, c_2=12, a_e=1.3, \theta_e = 4, c_3=6,  c_4=3,  a_i=2,  \theta_i = 1.5 \},
\end{split}
\end{equation}
for which, as the bifurcation diagram in Fig.~\ref{fig:2d_bifDiag} shows, the system \eqref{eq:WCsys} for $A=0$ displays a limit cycle denoted by $\Gamma_0$ for some $(P, Q)$ values. In particular, in this paper we choose $(P, Q) = (2.5, 0)$, so the unperturbed limit cycle is near a Hopf bifurcation.

Besides the Wilson-Cowan equations, we model the external input $p(t)$ to the excitatory population by means of a positive $T'$-periodic function. In this paper we have chosen:
\begin{equation}\label{eq:perturbationCosinus}
p(t) = 1 + \cos\left(\frac{2\pi t}{T'}\right).
\end{equation}

\section{Dynamical Analysis}\label{sec:dyn_anal}

In this Section we will define the stroboscopic map of system \eqref{eq:WCsys} and study the bifurcations of its fixed points as the amplitude and the frequency of the perturbation $p(t)$ in \eqref{eq:perturbationCosinus} are varied. In particular, we will focus on the study of the 1:1 and 1:2 phase-locked states, as we will see in Section~\ref{sec:section4}, they can serve to interpret several aspects of the CTC theory. 

\subsection{The stroboscopic map}\label{sec:strbrMap}

To study the $T'$-periodic system \eqref{eq:WCsys} we use the stroboscopic map defined by 
\begin{eqnarray}\label{eq:genericMap}
F_{A}:\mathbb{R}^{2} &\to& \mathbb{R}^{2}, \notag \\
x &\to& F_{A}(x) = \phi_A(t_0 + T'; t_0, x),
\end{eqnarray}
where $\phi_A(t; t_0, x)$ is the solution of~\eqref{eq:WCsys} such that $\phi_A(t_0; t_0, x) = x$. Calculations in this paper will always assume that $t_0$ = 0.

As it is well known, periodic orbits of system \eqref{eq:WCsys} are given by the fixed and periodic points of the stroboscopic map \eqref{eq:genericMap}, whereas the quasi-periodic solutions correspond to its invariant curves \cite{AP90}. More precisely, if $\gamma(t) = \phi_A(t; t_0, x)$ is a solution of system~\eqref{eq:WCsys} and $[F_A(x)]^q = x$, then $\phi_A(t_0 + qT'; t_0, x) = x$ and therefore $\gamma(t)$ is a periodic orbit of system \eqref{eq:WCsys} with period $qT'$. Analogously, if $\gamma(t) = \phi_A(t; t_0, x)$ is a periodic orbit of period $T$ of~\eqref{eq:WCsys} with $\frac{T'}{T}=\frac{p}{q}$,  $p,q \in \mathbb{N}$, then
\begin{equation}\label{eq:phaselocking}
[F_{A}(x)]^q = \phi_A(t_0 + qT'; t_0, x) = \phi_A(t_0 + pT; t_0, x) = x.
\end{equation} 
Otherwise, if $T'/T \in \mathbb{R} \setminus \mathbb{Q}$, then the iterates of $F_A$ fill densely an invariant curve denoted by $\Gamma_A$.

In the perturbed Wilson-Cowan model \eqref{eq:WCsys}, the relationship~\eqref{eq:phaselocking} indicates a \textit{p:q} phase locked state between the population and the perturbation. This means that the neuronal population variables $r_e$ and $r_i$ have completed \textit{p} revolutions in the same time that the perturbation $p(t)$ has completed \textit{q} revolutions.

\subsection{Dynamics of the stroboscopic map $F_A$}\label{sec:bifAnalisys}

\begin{figure}[H]
	\begin{minipage}[c]{0.7\textwidth}
		\includegraphics[width=1\linewidth]{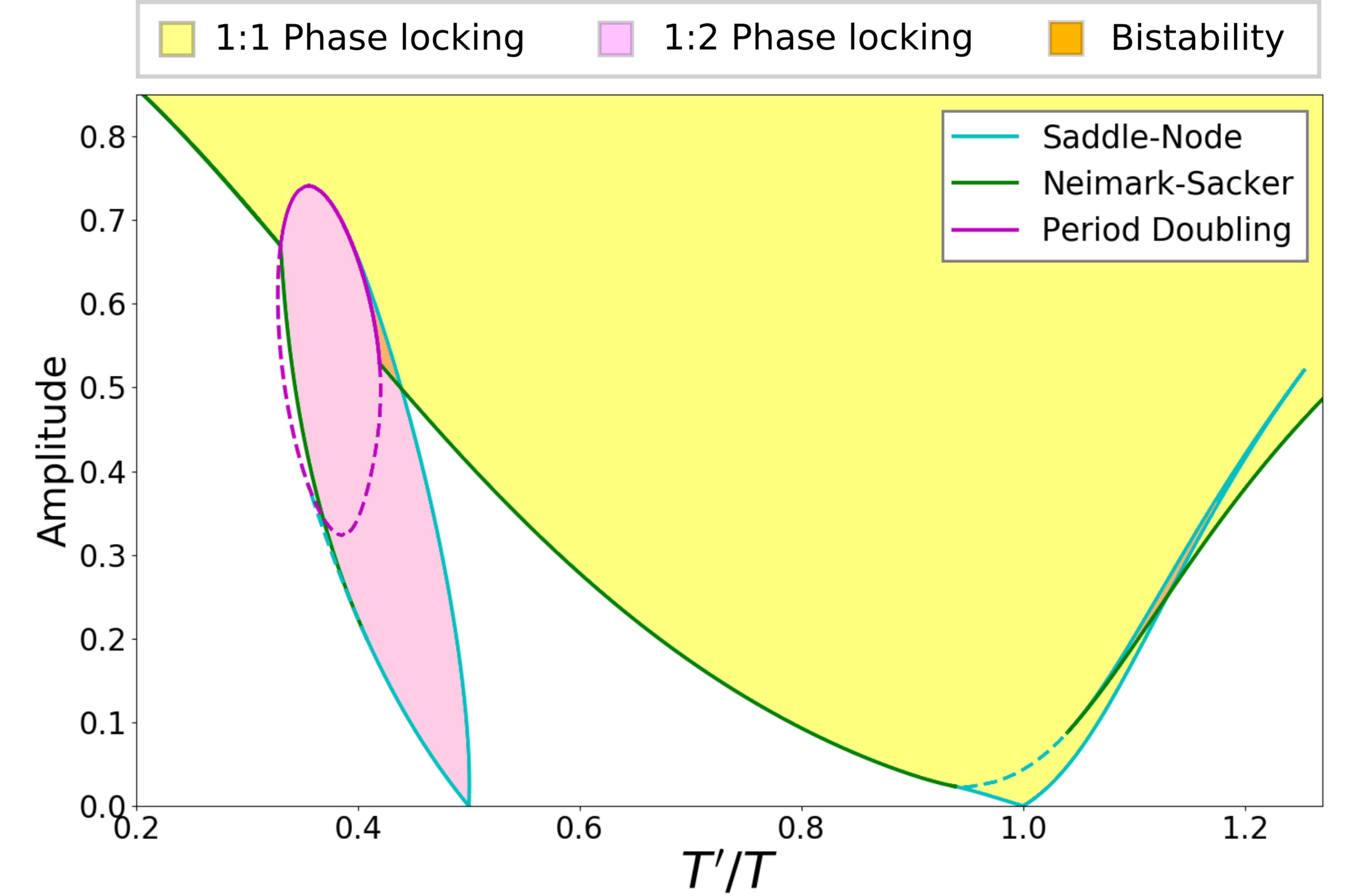}
	\end{minipage}\hfill
	\begin{minipage}[c]{0.27\textwidth}
		\caption{Bifurcation diagram for the fixed points of the stroboscopic map \eqref{eq:genericMap} of system \eqref{eq:WCsys} for the set of parameters $\mathcal{P}$ given in \eqref{eq:parametersChoice}, ($P$, $Q$) = (2.5, 0), as the frequency and the amplitude of the perturbation are varied. Solid curves correspond to bifurcations of stable fixed points whereas dashed curves correspond to bifurcations of unstable fixed points. The coloured regions correspond to different phase locking regimes: 1:1 phase-locking (yellow), 1:2 phase-locking (pink), bistability (orange). See text for more details.} \label{fig:HBbifDiagLarge}
	\end{minipage}
\end{figure}

Computing bifurcations of the fixed points of the stroboscopic map becomes relevant to identify different synchronous regimes as well as asynchronous ones.	Using the techniques described in Appendix \ref{sec:bifAnalysis} we can compute the bifurcation diagram for the fixed points of the stroboscopic map \eqref{eq:genericMap} of system \eqref{eq:WCsys} as the amplitude and the frequency of the perturbation $p(t)$ in \eqref{eq:perturbationCosinus} are varied. As Fig.~\ref{fig:HBbifDiagLarge} shows, the fixed points of the map $F_A$ undergo different bifurcations, namely, saddle-node, Neimark-Sacker and period doubling bifurcations, which bound the 1:1 and 1:2 phase locking areas, and allow for a natural identification of the synchronous regimes of interest. The yellow and pink regions correspond to 1:1 and 1:2 phase locked states of system \eqref{eq:WCsys}, respectively. We recall that they correspond to fixed points (1:1) and 2-periodic points (1:2) of the map $F_A$. The white regions may contain other \textit{p:q} phase-locked states, as well as asynchronous states. The orange regions contain more than one stable invariant object for the map $F_A$.

Next, we study in detail the dynamics predicted by the bifurcation diagram in Fig.~\ref{fig:HBbifDiagLarge}.  In particular, we will consider different $T'/T$ intervals  and study in detail the dynamics as the amplitude $A$ of the perturbation $p(t)$ is increased. We recall that by choosing the set of parameters $\mathcal{P}$ given in \eqref{eq:parametersChoice}, ($P$, $Q$) = (2.5, 0) and $A=0$, the phase space for system \eqref{eq:WCsys} shows the limit cycle $\Gamma_{0}$ and an unstable focus $P_1$ (see Fig.~\ref{fig:2d_bifDiag} bottom left). As both objects are normally hyperbolic we expect them to persist for weak enough amplitudes as an invariant curve $\Gamma_A$ and a fixed point $P_1$, respectively, for the corresponding stroboscopic map $F_A$ in \eqref{eq:genericMap}. The analysis that we perform focuses on 1:1 and 1:2 phase-locked states because they occupy the largest regions of the parameter space. Furthermore, as we will see 
in Section \ref{sec:section4}, the dynamics emerging in the 1:1 and 1:2 phase-locking regions can be interpreted in terms of the CTC theory which motivates this study. 
In particular, the 1:1 phase-locking pattern allows for the study of the modulation of the input gain whereas the 1:2 phase-locking pattern accounts for selective communication. 

\subsubsection*{Dynamics close to the saddle-node bifurcation at the 1:1 phase-locking region}

\begin{figure}[H]
	\begin{center}
		\includegraphics[width=160mm]{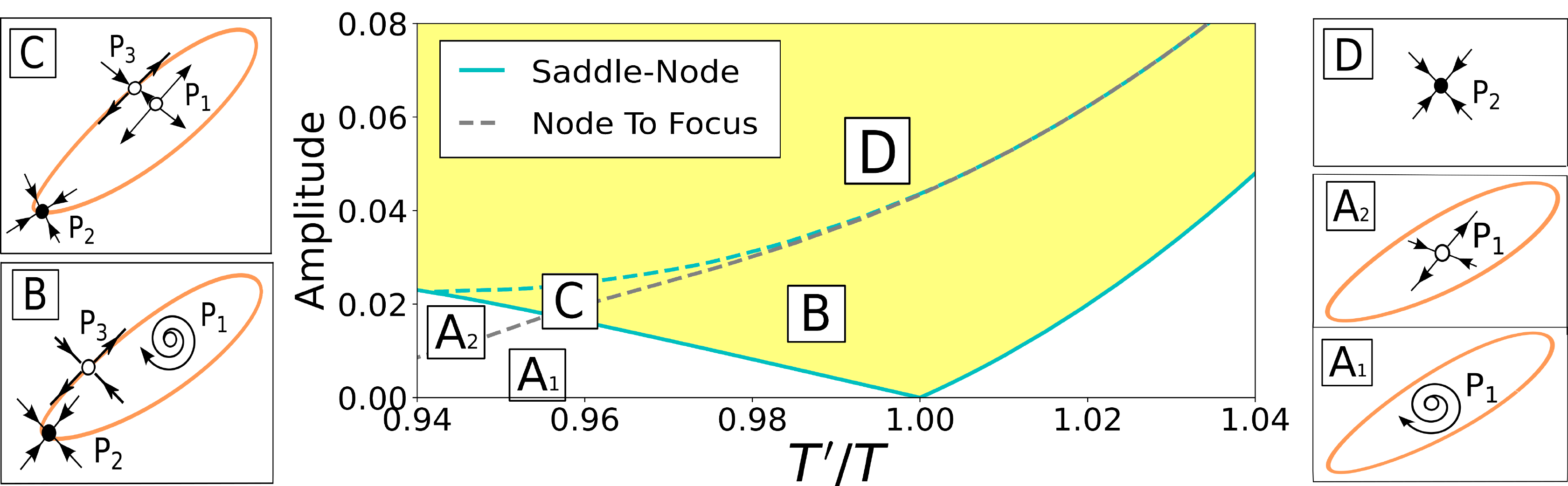} 
		\caption{Dynamics close to the saddle-node bifurcation at the 1:1 phase-locking region. Central panel shows a zoom of the bifurcation diagram for the map $F_A$ in Fig.~\ref{fig:HBbifDiagLarge} in the region close to the saddle-node bifurcation at the 1:1 phase-locking region. Panels A-D show a sketch of the phase space for the map $F_A$ in different parameter regions indicated accordingly in the central panel. Solid and empty dots correspond to stable and unstable fixed points, respectively, while orange curves correspond to a sketch of the invariant curves. Arrows indicate only the type of fixed point. See text for more details. }\label{fig:phaseSpaceSN}
	\end{center}
\end{figure}

The dynamics for values of $T'$ such that 0.9388 $< \frac{T'}{T} <$ 1.04, was studied in \cite{PerezCerveraHS17}. For the sake of completeness we recall here the main results, which are shown in Fig.~\ref{fig:phaseSpaceSN}. For $A$ small, an attracting invariant curve $\Gamma_A$ and an unstable focus $P_1$ inside it exist (regions $A_1$ and $B$). In region $A_1$ the invariant curve $\Gamma_A$ has no fixed points and once the saddle-node bifurcation (solid blue curve) is crossed (region B), there appear two fixed points on the invariant curve: a stable node $P_2$ and a saddle $P_3$, thus a SNIC bifurcation occurs, so $\Gamma_A$ consists of the union of the saddle $P_3$, its unstable invariant manifolds, and the stable node $P_2$. If the amplitude is increased (region C), $P_1$ becomes an unstable node (dashed gray curve). Furthermore, if the amplitude is increased further, $P_1$ coalesces with $P_3$ in an unstable saddle-node bifurcation (dashed blue curve), causing the disappearance of the invariant curve $\Gamma_A$ and the stable node $P_2$ remains as the unique fixed point (region D). Observe that it is possible to pass from region $A_1$ to region C, without passing through region B. In the region $A_2$ the unstable focus $P_1$ becomes an unstable node before crossing the saddle-node bifurcation curve (solid blue curve).

\subsubsection*{Dynamics close to the Neimark-Sacker bifurcation at the 1:1 phase-locking region}

\begin{figure}[H]
\begin{center}
\includegraphics[width=0.7\linewidth]{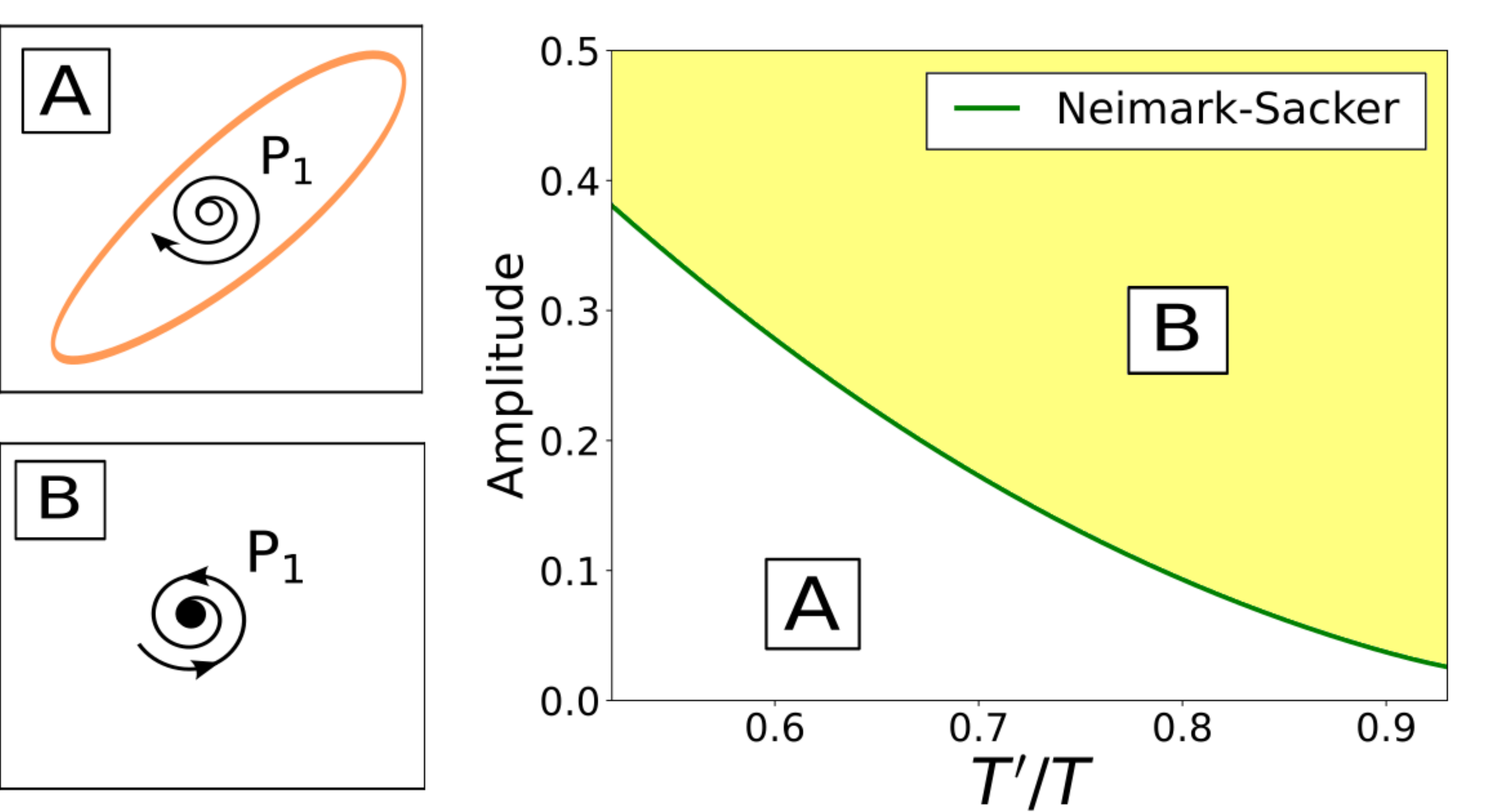} 
\end{center}

\caption{Dynamics close to the Neimark-Sacker bifurcation at the 1:1 phase-locking region. Right panel shows a zoom of the bifurcation diagram for the map $F_A$ in Fig.~\ref{fig:HBbifDiagLarge} in the region close to the Neimark-Sacker bifurcation at the 1:1 phase-locking region. Panels A-B show a sketch of the phase space for the map $F_A$ in different parameter regions indicated accordingly in the central panel. Solid and empty dots correspond to stable and unstable fixed points, respectively, while orange curves correspond to a sketch of the invariant curves. Arrows indicate only the type of fixed point. See text for more details. } \label{fig:phaseSpaceNS}
\end{figure}

The dynamics for values of $T'$ such that 0.51 $< \frac{T'}{T} <$ 0.9388, was studied in \cite{PerezCerveraHS17}. For the sake of completeness we recall here the main results, which are shown in Fig.~\ref{fig:phaseSpaceNS}. For $A$ small, the attracting invariant curve $\Gamma_A$ has no fixed points of $F_A$, and an unstable focus $P_1$ exists inside $\Gamma_A$ (region A). If the amplitude $A$ is further increased, a Neimark-Sacker bifurcation occurs (green curve). At this point, the curve $\Gamma_A$ collapses to $P_1$ and disappears, while $P_1$ becomes a stable focus (region B). 

\subsubsection*{Dynamics on the left hand side of the 1:2 phase-locking region}

\begin{figure}
	\centering
	\includegraphics[width=1\linewidth]{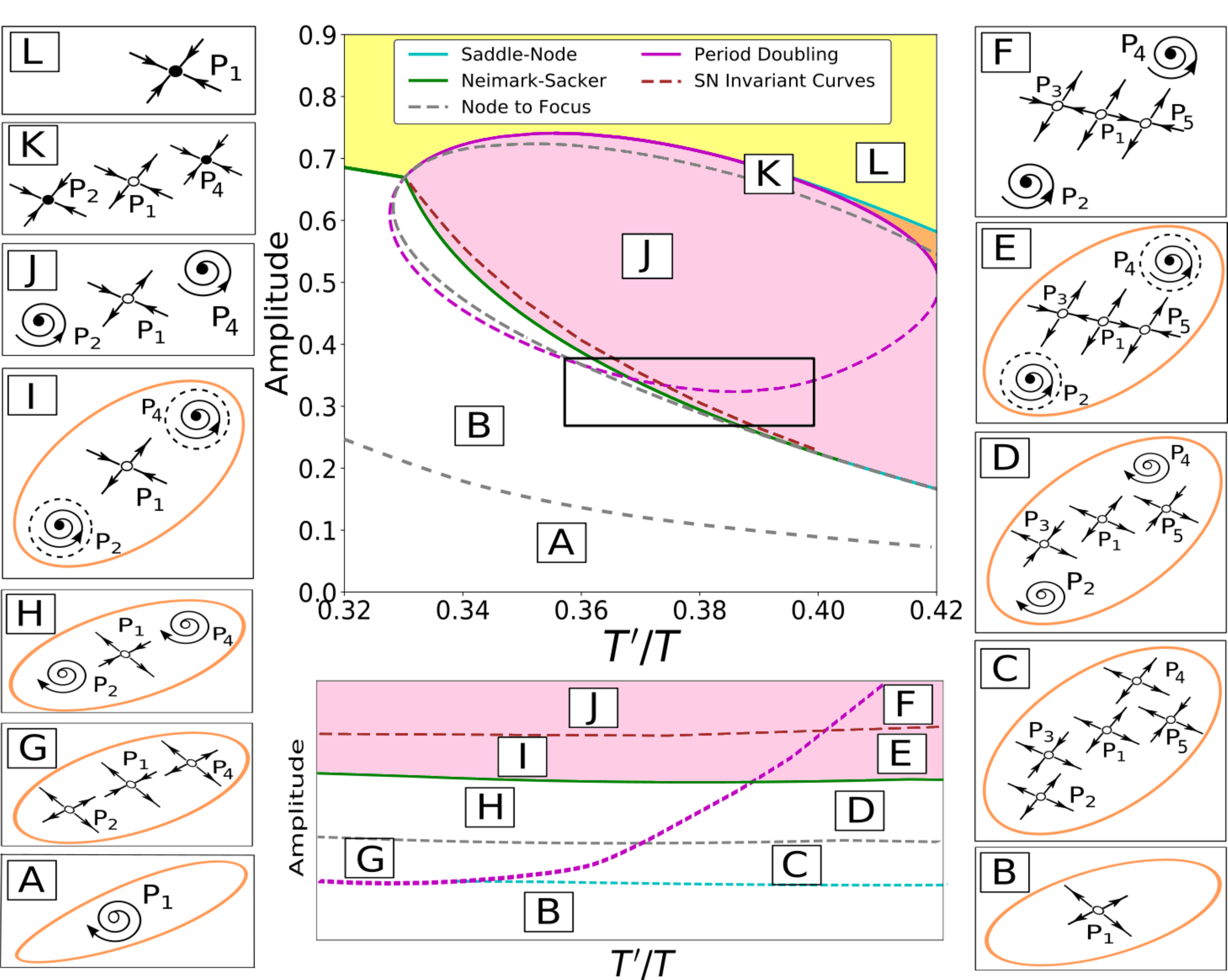} 
	\caption{Dynamics close to the left hand side of the 1:2 phase-locking region. Top central panel shows a zoom of the bifurcation diagram for the map $F^2_A$ in Fig.~\ref{fig:HBbifDiagLarge} for $0.32 < T'/T < 0.42$. The bottom central panel contains a sketch of the bifurcations inside the black rectangle. Panels A-L show a sketch of the phase space for the map $F^2_A$ in different parameter regions indicated accordingly in the central panel. Solid and empty dots correspond to stable and unstable fixed points, respectively, while orange curves correspond to a sketch of the invariant curves. Arrows indicate only the type of fixed point. See text for more details.}
	\label{fig:areaA5}
\end{figure}

For values of $T'$ such that 0.32 $< \frac{T'}{T} <$ 0.42, the phase portrait for the map $F_A$ in different regions of the parameter space is shown in Fig.~\ref{fig:areaA5}. For $A$ small, there exists a stable invariant curve and an unstable focus $P_1$  (region A). When the amplitude increases, $P_1$ becomes an unstable node when it crosses the dashed grey curve (region B). If the amplitude increases more, one finds, depending on the $T'$ value considered, different bifurcation curves where there appear unstable fixed points for the map $F^2_{A}$. Next, we describe the dynamics in the zoomed region in Fig.~\ref{fig:areaA5} containing a sketch of such bifurcations.
For 0.355 $< \frac{T'}{T} <$ 0.4, as $A$ increases an unstable saddle-node bifurcation for $F^2_{A}$ is crossed (dashed blue curve) and two saddles ($P_3$ and $P_5$) and two unstable nodes ($P_2$ and $P_4$) appear as fixed points for the map $F^2_{A}$  (region C). By contrast for 0.32 $< \frac{T'}{T} <$ 0.355, one finds a period doubling bifurcation (dashed purple curve), at which there appear two unstable nodes ($P_2$ and $P_4$) and the unstable node $P_1$ becomes a saddle (region G). In both cases, a slight increase of the amplitude $A$ causes the unstable nodes $P_2$ and $P_4$ of $F^2_{A}$ to become unstable focuses at the dashed grey line (regions D and H). These focuses change their stability at a subcritical Neimark-Sacker bifurcation (green curve). Thus, an unstable invariant curve appears surrounding each of the stable focuses (regions E and I, respectively), generating a situation of bistability between the invariant curve $\Gamma_A$ and the fixed points $P_2$ and $P_4$. 
These unstable invariant curves undergo a homoclinic bifurcation (not shown) generating a unique unstable invariant curve which collides with the stable invariant curve $\Gamma_A$ at a saddle-node bifurcation of invariant curves (brown dashed curve). Therefore, the stable focuses $P_2$ and $P_4$ remain as the unique attractors (regions F and J). Additionally, the saddles $P_3$ and $P_5$ in region F disappear at a period doubling bifurcation (dashed purple curve) and transitioning from region F to J. 

So, as the amplitude $A$ increases, by means of different bifurcation routes, that depend on $T'/T$, the map $F^2_{A}$ shows the phase portrait depicted in region J, consisting of a saddle $P_1$ and two stable focuses $P_2$ and $P_4$. Finally, for large enough amplitudes, the stable focuses $P_2$ and $P_4$ for the map $F^2_{A}$ become stable nodes when crossing the dashed grey line (region K). Increasing the amplitude further, both points collapse at a period doubling bifurcation (solid purple line) where the saddle $P_1$ becomes a stable node (region L).

\subsubsection*{Dynamics on the right hand side of the 1:2 phase-locking region}

\begin{figure}[H]
	\centering
	\includegraphics[width=1\linewidth]{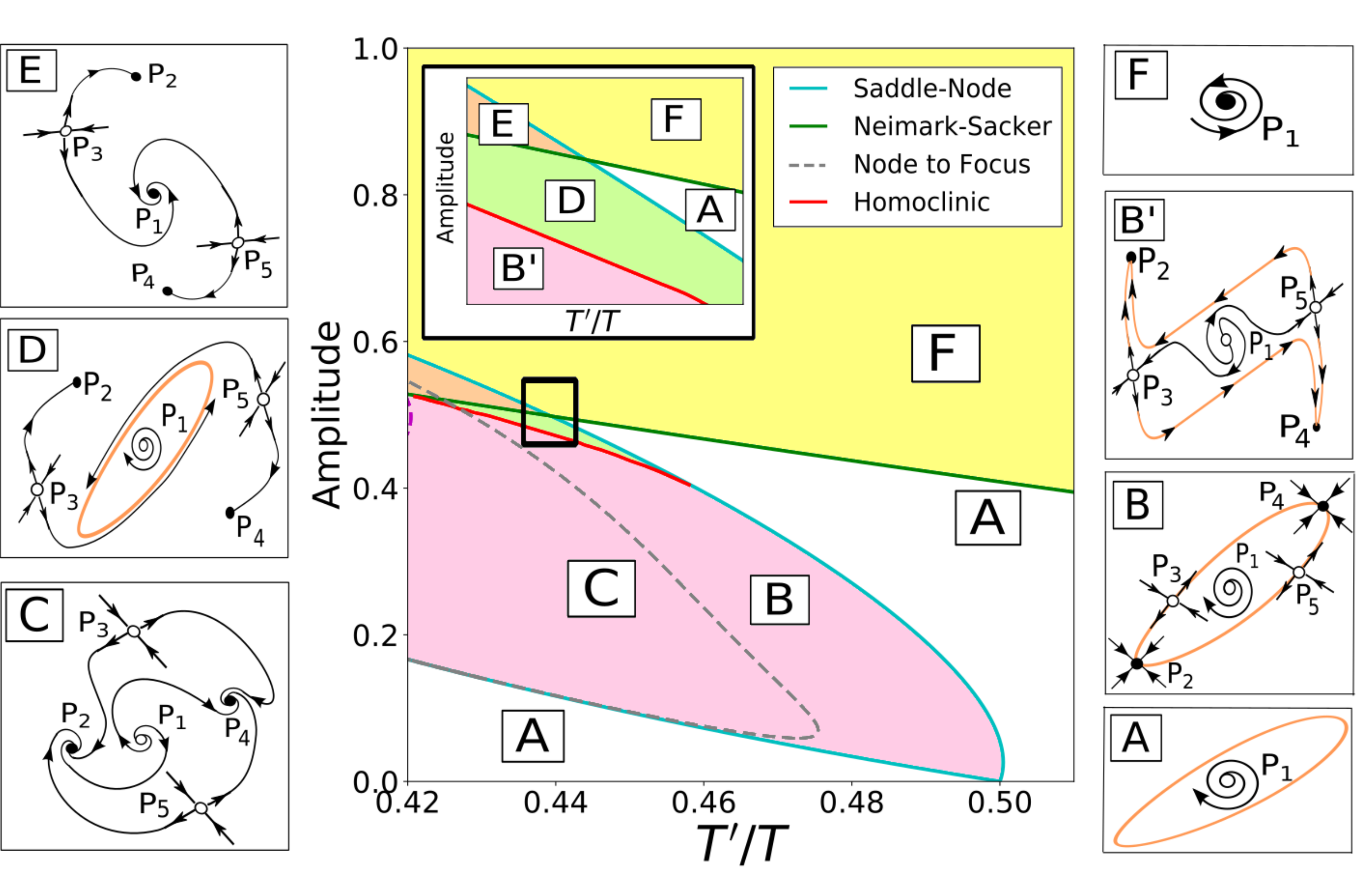} 
	\caption{Dynamics close to the right hand side of the 1:2 phase-locking region. Central panel shows a zoom of the bifurcation diagram for the map $F^2_A$ in Fig.~\ref{fig:HBbifDiagLarge} for $0.42 < T'/T < 0.51$. The inset panel contains a sketch of the bifurcations inside the black rectangle. Panels A-F show a sketch of the phase space for the map $F^2_A$ in different parameter regions indicated accordingly in the central panel. Solid and empty dots correspond to stable and unstable fixed points, respectively, while orange curves correspond to a sketch of the invariant curves. Arrows indicate the type of fixed point in panels A, B and F while for the rest we include an sketch of the globalization of the invariant manifolds. See text for more details.}
	\label{fig:areaA4}
\end{figure}

For values of $T'$ such that 0.42 $< \frac{T'}{T} <$ 0.51, the phase portrait for the map \eqref{eq:genericMap} in different regions of the parameter space is shown in Fig.~\ref{fig:areaA4}. For $A$ small, the attracting invariant curve $\Gamma_A$ generated from the unperturbed limit cycle $\Gamma_{0}$ has no fixed points, and an unstable focus $P_1$ exists inside $\Gamma_A$ (region A). When the amplitude is increased, a saddle node bifurcation curve of $F^2_A$ is crossed (blue curve), and there appear four fixed points on the invariant curve for the map $F_{A}^2$: two stable nodes ($P_2$ and $P_4$), and two saddles ($P_3$ and $P_5$) (region B). The invariant curve consists of the union of both saddles and their unstable invariant manifolds with the fixed points $P_2$ and $P_4$. 

For values of $T'$ such that 0.48 $< \frac{T'}{T} <$ 0.51, as the amplitude is increased, $P_2$, $P_3$, $P_4$ and $P_5$ pair-collide again on a saddle-node bifurcation (blue curve) and disappear leaving an attracting invariant curve without periodic points and the unstable fixed point $P_1$ (region A). The amplitude of this invariant curve decreases as the amplitude $A$ increases until it reaches a supercritical Neimark-Sacker bifurcation (green curve) for the map $F_{A}$ leaving just a stable focus as the unique fixed point $P_1$ (Region F).

For values of $T'$ such that 0.42 $< \frac{T'}{T} <$ 0.48, as the amplitude $A$ increases, very close to the saddle node bifurcation curve, $P_2$ and $P_4$ become stable focuses at the grey dashed curve (region C). In this region the only stable objects are the focuses $P_2$ and $P_4$. As the amplitude is increased further, this situation is maintained until $P_2$ and $P_4$ cross again the grey dashed curve and become stable nodes, and the invariant curve is the union of the saddle points $P_3$ and $P_5$ and their unstable invariant manifolds with the fixed points $P_2$ and $P_4$ (region B').

As the amplitude is increased further, a homoclinic bifurcation is crossed (red curve) and an invariant curve appears (region D). Therefore, we have found a region where our system presents bistability between an attracting invariant curve and the fixed points $P_2$ and $P_4$ for $F^2_A$. If the amplitude is increased further, a Neimark-Sacker bifurcation is crossed (green curve), thus the invariant curve disappears and $P_1$ changes stability (region E). This situation of bistability between a 2-periodic orbit and a fixed point $P_1$ of the map $F_A$ persists as the amplitude increases further until it reaches a saddle-node bifurcation (blue curve) for $F^2_{A}$ when the stable focus $P_1$ remains as the only fixed point (region F). See Appendix \ref{sec:bifAnalysis} for the description of the procedure followed to compute the homoclinic bifurcation curve.

\subsubsection*{Dynamics on the bottom right of the 1:1 phase-locking region}

\begin{figure}[H]
	\centering
	\includegraphics[width=1\linewidth]{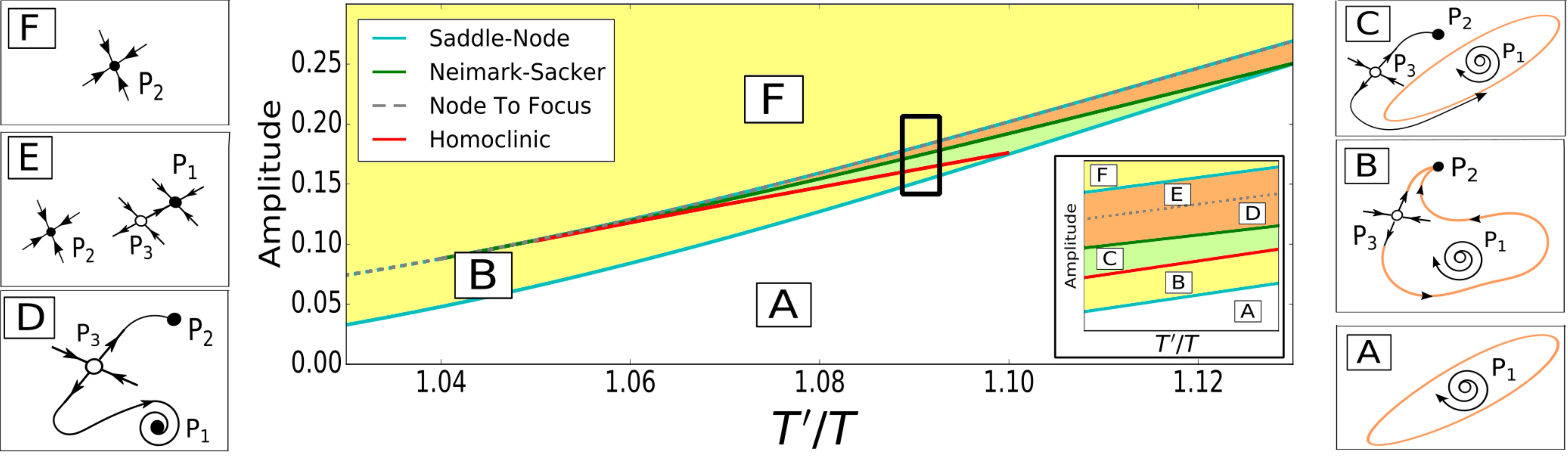} 
	\caption{Dynamics close to the bottom right of the 1:1 phase-locking region. Central panel shows a zoom of the bifurcation diagram for the map $F_A$ in Fig.~\ref{fig:HBbifDiagLarge} for $1.04 < T'/T < 1.125$. The inset  contains a sketch of the bifurcations inside the black rectangle. Panels A-F show a sketch of the phase space for the map $F_A$ in different parameter regions indicated accordingly in the central panel. Solid and empty dots correspond to stable and unstable fixed points, respectively, while orange curves correspond to a sketch of the invariant curves. Arrows indicate only the type of fixed point. See text for more details.}
	\label{fig:areaA6}
\end{figure}

For values of $T'$ such that 1.04 $< \frac{T'}{T} <$ 1.125, the phase portrait for the map \eqref{eq:genericMap} in different regions of the parameter space can be seen in Fig.~\ref{fig:areaA6}. The invariant curve $\Gamma_A$ (region A) evolves as the amplitude increases until crossing a saddle-node bifurcation (blue curve). At this bifurcation a stable node $P_2$ and a saddle $P_3$ are born and the invariant curve consists of the union of the saddle $P_3$ and its unstable invariant manifolds with the stable node $P_2$ (region B). Increasing the amplitude, a homoclinic bifurcation is crossed (red curve) and there appears a stable invariant curve without fixed points generating bistability between the invariant curve itself and the fixed point $P_2$ (region C). This invariant curve collapses at a Neimark-Sacker bifurcation (green curve), where the focus becomes stable, generating bistability between the fixed points $P_1$ and $P_2$ (region D). This situation persists until $P_1$ becomes a stable node at the grey dashed line (region E) which coalesces with the saddle $P_3$ at a saddle node bifurcation (blue curve) and disappears leaving $P_2$ as the unique (stable) fixed point (region F). See Appendix \ref{sec:bifAnalysis} for the description of the procedure followed to compute the homoclinic bifurcation curve.

\subsubsection*{Dynamics on the top right of the 1:1 phase-locking region}

\begin{figure}[H]
	\centering
	\includegraphics[width=1\linewidth]{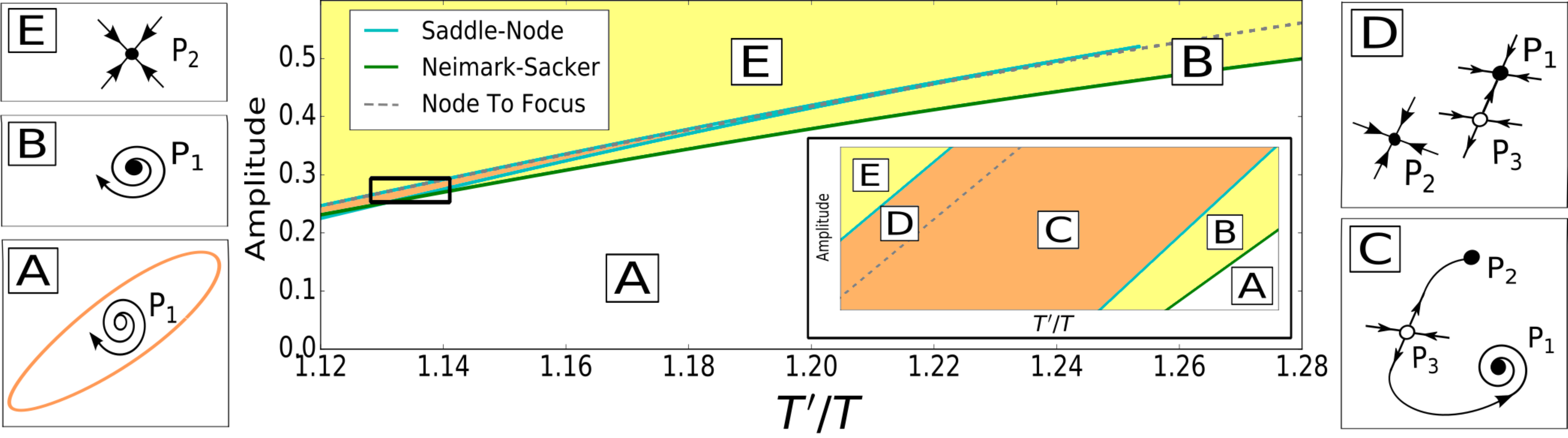} 
	\caption{Dynamics close to the top right of the 1:1 phase-locking region. Central panel shows a zoom of the bifurcation diagram for the map $F_A$ in Fig.~\ref{fig:HBbifDiagLarge} for $1.125 < T'/T < 1.255$. The inset contains a sketch of the bifurcations inside the black rectangle. Panels A-F show a sketch of the phase space for the map $F_A$ in different parameter regions indicated accordingly in the central panel. Solid and empty dots correspond to stable and unstable fixed points, respectively, while orange curves correspond to a sketch of the invariant curves. Arrows indicate only the type of fixed point. See text for more details.}
	\label{fig:areaA7}
\end{figure}

For values of $T'$ such that 1.125 $< \frac{T'}{T} <$ 1.255, the phase portrait for the map \eqref{eq:genericMap} in different regions of the parameter space can be seen in Fig.~\ref{fig:areaA7}. As the amplitude is increased, the invariant curve $\Gamma_A$ (region A) collapses at a Neimark-Sacker bifurcation (green curve), where the stability of the focus $P_1$ changes (region B). If the amplitude is increased further, a stable node $P_2$ and a saddle $P_3$ appear at a saddle-node bifurcation (blue curve), generating a situation of bistability between the focus $P_1$ and the node $P_2$ (region C). As the amplitude $A$ increases further, $P_1$ becomes a stable node at the grey dashed line (region D) and coalesces with $P_3$ at a saddle-node bifurcation, leaving node $P_2$ as the unique (stable) fixed point (region E).

\section{Implications for CTC Theory}\label{sec:section4}

In this section, we interpret the results obtained in Section~\ref{sec:dyn_anal} in terms of the CTC framework. More precisely, we explore the implications of the 1:1 phase-locked states in the modulation of the input gain, the 1:2 phase-locked states in selective communication and the bistability regions in regulating communication.

According to the CTC theory, phase locking between the emitting and receiving populations is required to establish an effective communication. Nevertheless, as an effective communication is characterized by a noticeable increase of the response of the excitatory receiving population, it turns out that the timing between the input and the inhibitory response might modulate the response of the excitatory receiving population. Indeed, inputs preceding inhibition may participate effectively in the response of the receiving population, thus increasing it. By contrast, inputs following the inhibitory action may be partially or totally silenced and, thus, have almost no noticeable effect in the receiving population. Next, in order to explore the relationship between the timing of the inhibition and the increase of the response of the excitatory population to the input we define and compute two magnitudes, $\Delta \theta$ and $\Delta \alpha$, on the phase-locking areas of interest.
%we compute two features for the fixed points of the map $F_A$ into the phase-locked areas. 
%As an effective communication implicates a noticeable increase of the response of the receiving population we will compute 
%the phase difference between the input inhibitory population, respectively. 

\begin{figure}
	\centering
	\includegraphics[width=1\linewidth]{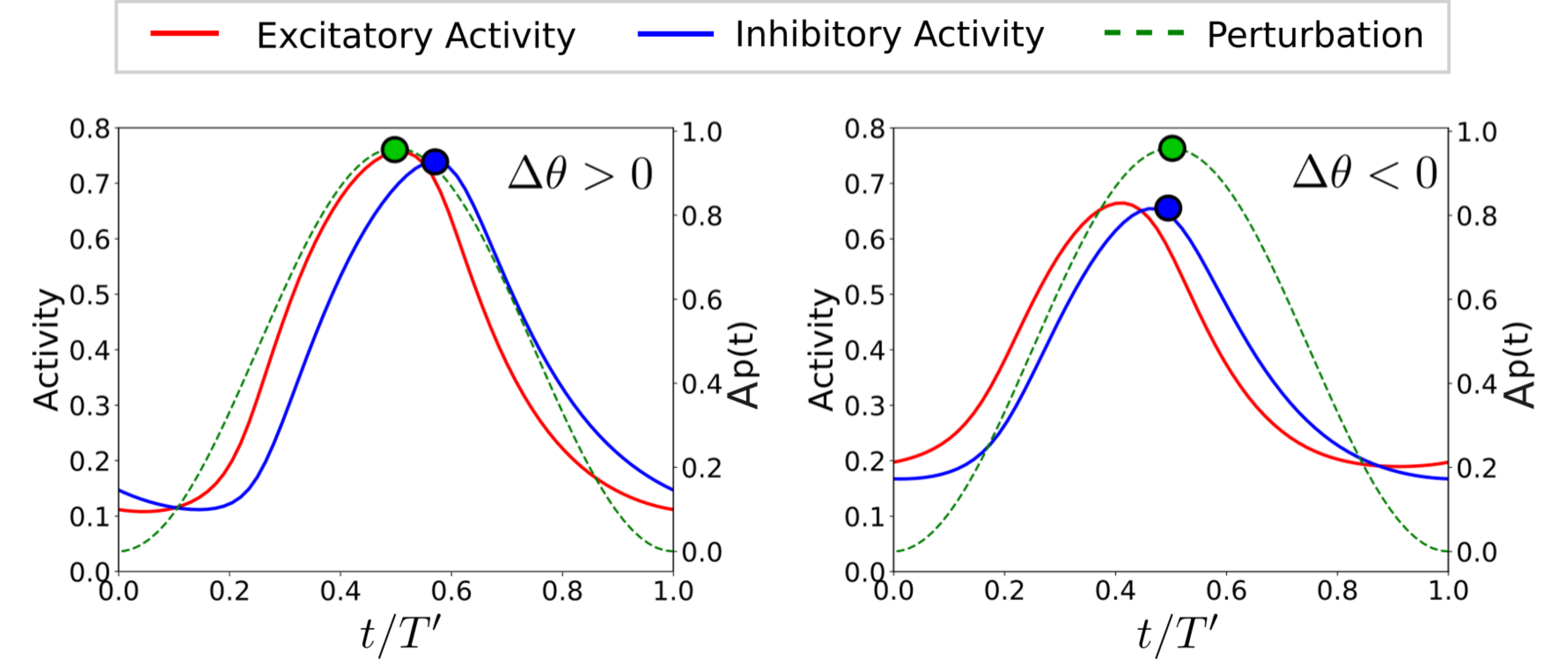} 
	\caption{Effect of the phase $\Delta \theta$ defined in Eq.~\eqref{eq:timeDifference11} onto the excitatory response. For $\Delta \theta > 0$ (left) the input (green dashed curve) precedes the inhibition (blue curve), whereas for $\Delta \theta < 0$ (right) the input follows the inhibition. Parameters used are $(A, T'/T) = (0.47, 1.2)$ (left) and $(A, T'/T) = (0.47, 1.24)$ (right).} Notice how, for two perturbations of identical amplitude and similar frequency, changes on the sign of $\Delta \theta$ imply a change on the excitatory population response (red curve).
	\label{fig:implications11}
\end{figure}

In particular, we define $\Delta \theta$ to compute
the phase difference between the maximum of the inhibitory population and the maximum of the perturbation, that is,
\begin{equation}\label{eq:timeDifference11}
\Delta \theta = \frac{t_{inh} - t_{pert}}{T'}, \quad \quad \Delta \theta \in [-0.5, 0.5),
\end{equation}
where $t_{inh}$ and $t_{pert}$ are the times at which $I(t)$ and $p(t)$ of system \eqref{eq:WCsys} achieve a maximum inside a cycle.  
Notice that when $\Delta \theta$ is positive, the perturbation precedes the activation of the inhibitory population, so we expect that the excitatory receiving population is sensitive to the input. On the contrary, when $\Delta \theta$ is negative, the perturbation follows the activation of the inhibitory population, so we expect that the excitatory receiving population is less sensitive to the input due to the presence of inhibition (see Fig.~\ref{fig:implications11}).

In addition, $\Delta \alpha$ computes the maximum of the activity of the excitatory population $\alpha_A$, normalized by the maximum of the activity of the unperturbed excitatory population $\alpha_0$, that is,
\begin{equation}\label{eq:amplitudeDifference11}
\Delta \alpha = \frac{\alpha_A}{\alpha_0}.
\end{equation}
Notice that when the rate $\Delta \alpha$ is greater than one, the perturbation increases the amplitude of $r_e(t)$. Therefore, the larger $\Delta \alpha$ the more effective the input. 

Next, we compute both magnitudes $\Delta \theta$ and $\Delta \alpha$ for each of the two phase locked regions considered. Notice that results for both regions can be interpreted differently. In the 1:1 region, as there is just one input per period, the input can precede or follow the inhibitory action, whereas in the 1:2 case, as there are two inputs per period, we expect one to precede and the other to follow the inhibitory action.  %and discuss its implications for the CTC framework. Furthermore, our results in Fig. X consider two different phase-locking areas 1:1 y la 1:2 from which we can extract different conclusions. More precisely, 

\subsection{Modulation of the Input Gain (1:1 phase-locking region)}\label{sec:sec41} 

\begin{figure}
	\centering
	\includegraphics[width=1\linewidth]{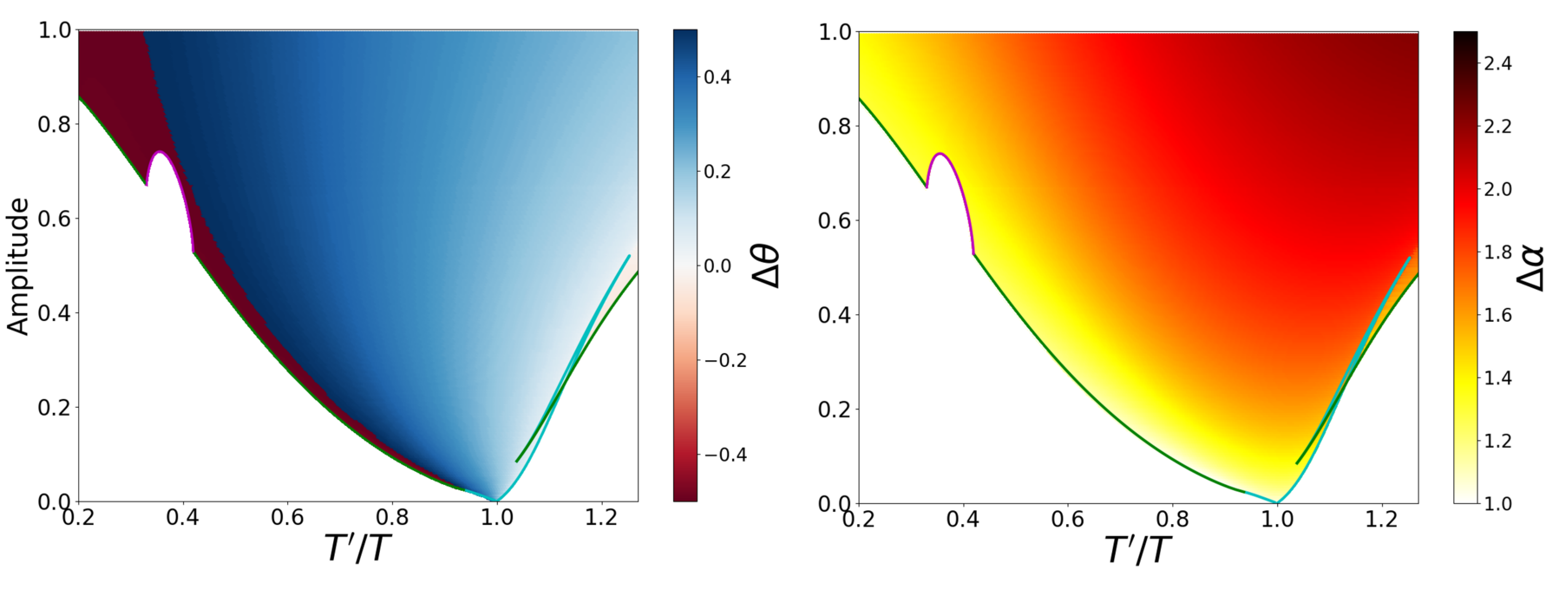} 
	\caption{For the 1:1 phase-locking region we show the phase difference $\Delta \theta$ defined in Eq.~\eqref{eq:timeDifference11} (left) 
	and the amplitude increase factor $\Delta \alpha$ defined in eq.~\eqref{eq:amplitudeDifference11} (right).}
	\label{fig:phaseAmplitudes11}
\end{figure}

By looking at the bifurcation diagram in Fig.~\ref{fig:HBbifDiagLarge}, we observe that there is a large region of 1:1 entrainment, which corresponds to the yellow region. To investigate the features 
of this entrainment we compute the quantities $\Delta \theta$ and $\Delta \alpha$ described above (see Fig.~\ref{fig:phaseAmplitudes11}).

We observe that, in general, $0 \leq \Delta \theta \leq 0.5$, indicating that inhibition typically follows the input. Predominance of positive values of $\Delta \theta$ (blue in the left panel of Fig.~\ref{fig:phaseAmplitudes11}) seems to indicate that the perturbation will have a positive effect onto the activity of the excitatory population. Indeed, we observe that the activity of the excitatory population increases since $\Delta \alpha > 1$ (see the right panel for Fig.~\ref{fig:phaseAmplitudes11}). Nevertheless, this increase is not the same for all the points in the 1:1 phase-locking region. Notice that, as it is expected, the response of the receiving population is larger as the amplitude of the input increases. Nevertheless, for a fixed forcing amplitude $A$, the factor $\Delta \alpha$ is lower near the borders of the 1:1 phase-locking region (white and red regions), where the inhibition action precedes the input ($\Delta \theta<0$) and it can suppress totally or partially the input effect.

In conclusion, the 1:1 phase-locking pattern naturally produces a stable phase relationship that is optimal for CTC in the sense that it promotes an increase in the firing rate activity of the receiving population. Interestingly, near the boundaries of the 1:1 region this situation is reversed ($\Delta \theta < 0$) so the perturbation follows the inhibitory action.

\subsection{Selective communication (1:2 phase-locking region)} 

\begin{figure}[H]
	\begin{minipage}[c]{0.65\textwidth}
		\includegraphics[width=1\linewidth]{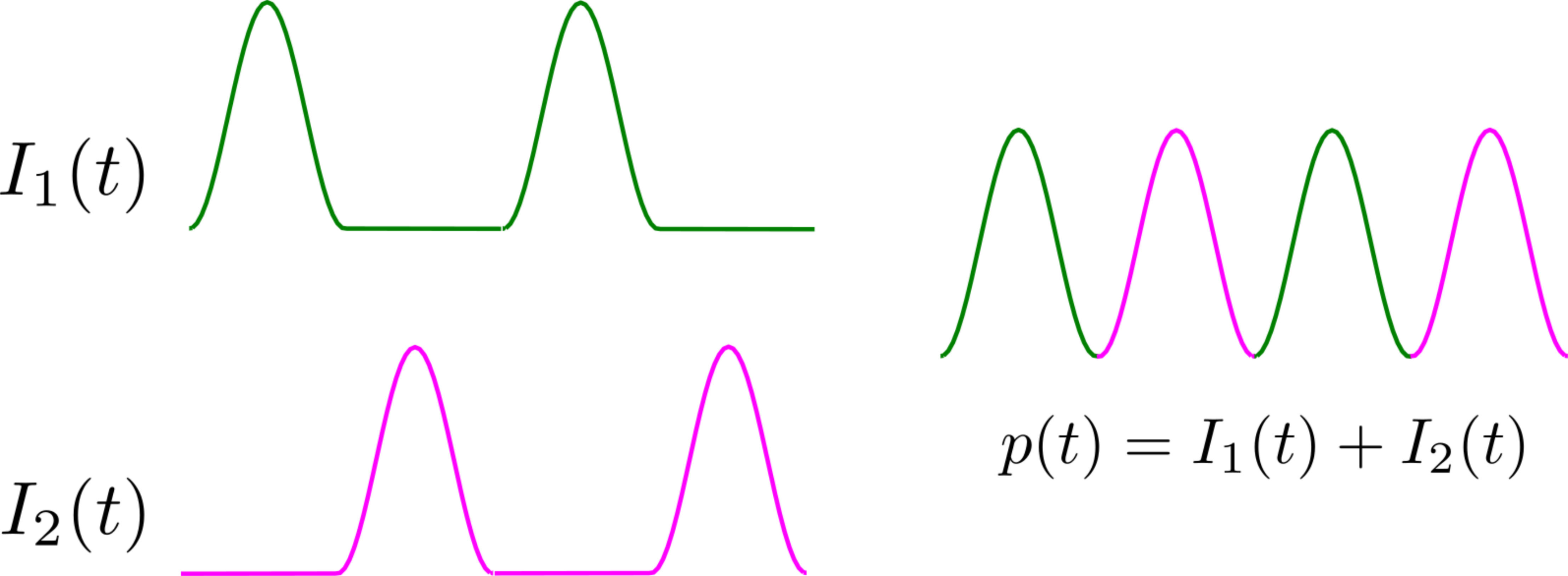} 
	\end{minipage}\hfill
	\begin{minipage}[c]{0.28\textwidth}
		%\caption{\textcolor{red}{The input in the 1:2 phase-locking region is interpreted as the sum of two identical inputs $I_1(t)$ and $I_2(t)$ in anti-phase coming from different sources, so $p(t) = I_1(t) + I_2(t)$, in order to study selective communication. See text for more details} } \label{fig:ctcFramework_v2}
		The 1:2 phase locked states can account for selective communication as the input $p(t)$ can be interpreted as the sum of two identical competing inputs $I_1(t)$ and $I_2(t)$ in anti-phase coming from different sources. See text for more details.
		\label{fig:ctcFramework_v2}
	\end{minipage}
\end{figure}

In the previous Section, we have shown that for the 1:1 phase-locking region, the input typically precedes the inhibitory response. 
This is especially interesting when studying the forcing with higher frequencies, as it is the case of the 1:2 phase-locking region. In this region, the input undergoes two cycles for one cycle of activity of the receiving population. Because of this, we expect that one of the input cycles precedes the inhibitory action whereas the other one follows it. Indeed, we can interpret the input in the 1:2 phase-locking region as two identical inputs, $I_1(t)$ and $I_2(t)$, from two different emitting neural populations, which  arrive to the receiving population separated by a half-period (see Fig.~\ref{fig:ctcFramework_v2}), and study competition between inputs. That is, we explore whether one input phase-locks at an optimal phase so that it increases the post-synaptic response, while the other one is ignored \cite{Borgers05, borgers2008gammaB}. In the CTC context, this situation is known as selective communication (only one pre-synaptic population communicates effectively) \cite{fries2005}. %As one input precedes the inhibition, whereas the other one follows it, the 1:2 phase-locking region accounts for an interpretation in terms of selective communication in the context of CTC theory. 
We refer to input $I_1(t)$ as the one that precedes the main inhibitory response and $I_2(t)$ as the one that follows it. Thus, we expect that input $I_1$ produces an increase in the activity of the receiving population whereas the other one $I_2$, is ignored. 

Similarly as the procedure followed in the 1:1 phase-locking case, we will study selective communication by computing  the timing between the input and the inhibitory response $\Delta \theta$ and the rate change in the response of the excitatory receiving population $\Delta \alpha$. Since in this case the input is interpreted as the sum of two inputs (Fig.~\ref{fig:ctcFramework_v2}), we will compute $\Delta \theta$ and $\Delta \alpha$ for each input, provided that each input generates a response of the population. For the first input $I_1(t)$, as we assume that it will always precede the inhibitory response, we can always compute the following magnitudes
\begin{equation}\label{eq:firstInputMagnitudes}
\Delta \theta_1 = \frac{t^{(1)}_{inh} - t^{(1)}_{pert}}{T'}, \quad \quad \quad \Delta \alpha_1 = \frac{\alpha^{(1)}_A}{\alpha_0},
\end{equation}
where $t^{(1)}_{inh}$ is the time at which $r_i(t)$ achieves a maximum inside a cycle and $t^{(1)}_{pert}$ is the time at which $I_1(t)$ achieves a maximum inside the interval $0 < t < T'$. Moreover, we denote by $\alpha^{(1)}_A$ the value of the excitatory activity at the main maximum.
 
\begin{figure}[H]
	\centering
	\includegraphics[width=1\linewidth]{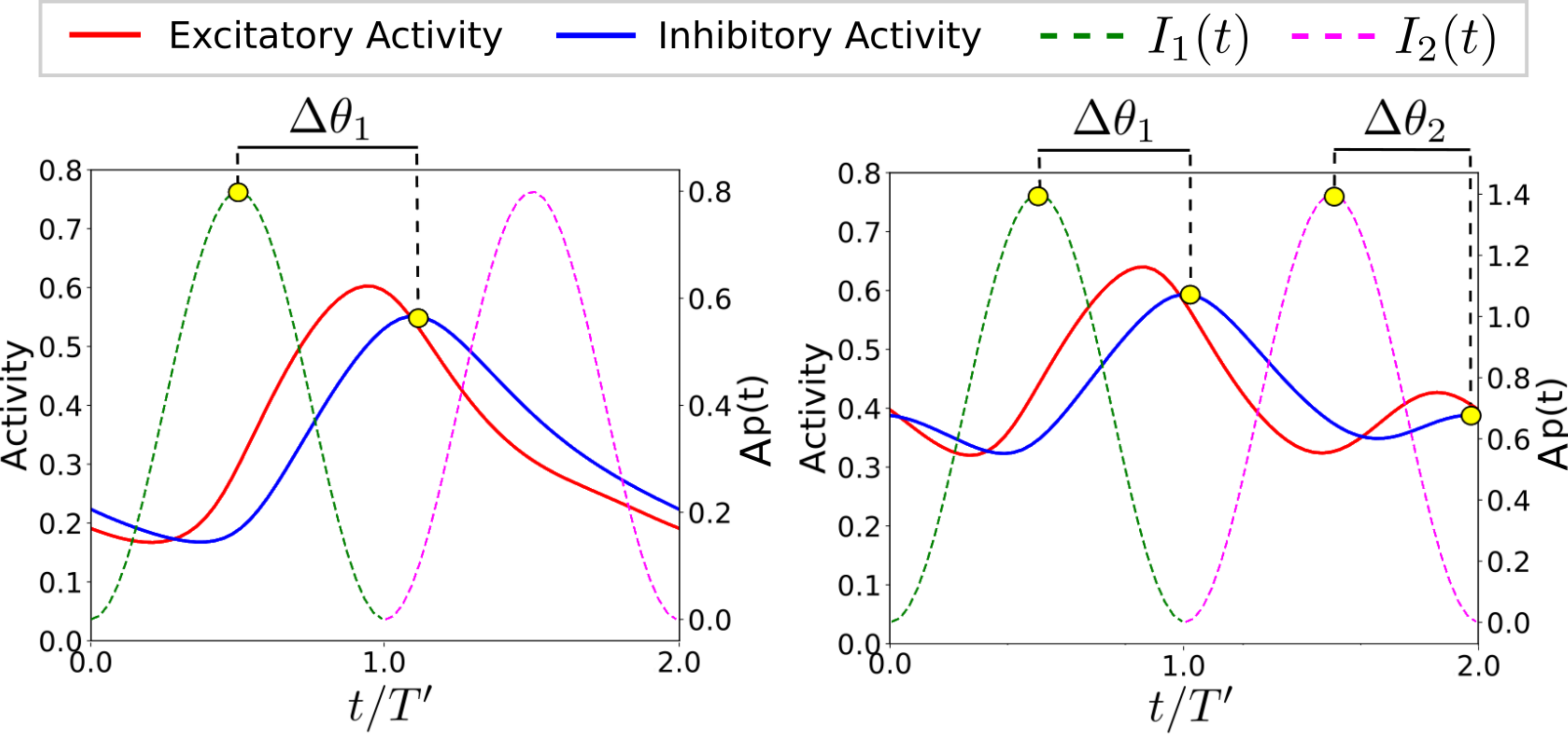} 
	\caption{Stable solution for system \eqref{eq:WCsys} with forcing parameters $(A, T'/T) = (0.4, 0.4)$ (left) and $(A, T'/T) = (0.7, 0.38)$ (right). In the left panel, the first input (green curve) elicits a response of the excitatory population (red curve), whereas the second input (purple curve) is not. By contrast, in the right panels, both inputs elicit a response (two bumps in the excitatory and inhibitory  activity). In the left panel we compute only $\Delta \theta_1$ and in the right panel we compute $\Delta \theta_1$ and $\Delta \theta_2$.}
	\label{fig:implications12}
\end{figure}

Nevertheless, for the second input $I_2(t)$ the situation is not so straightforward. We will consider that this input elicits a response from the receiving population if the activity of the excitatory/inhibitory population shows, apart from the main maximum, a second peak (see Fig.~\ref{fig:implications12}). In that case, we will also compute
\begin{equation}\label{eq:secondInputMagnitudes}
\Delta \theta_2 = \frac{t^{(2)}_{inh} - t^{(2)}_{pert}}{T'}, \quad \quad \quad \Delta \alpha_2 = \frac{\alpha^{(2)}_A}{\alpha_0},
\end{equation}
where $t^{(2)}_{inh}$ is the time at which $r_i(t)$ achieves a second local maximum inside a cycle and $t^{(2)}_{pert}$ is the time at which $I_2(t)$ achieves a maximum inside a cycle, so $T' < t^{(2)}_{pert} < 2T'$. Moreover, we denote by $\alpha^{(2)}_A$ the value of the excitatory activity at the second local maximum.

We remark that, differently from the 1:1 case in which the perturbation can either follow or precede the inhibition, so $-0.5 < \Delta \theta < 0.5$, in this case both magnitudes $\Delta \theta_1$ and $\Delta \theta_2$ are defined in such a way that they always precede an inhibitory response, so $\Delta \theta_1, \Delta \theta_2 \in [0, 1]$.

\begin{figure}
	\centering
	\includegraphics[width=1\linewidth]{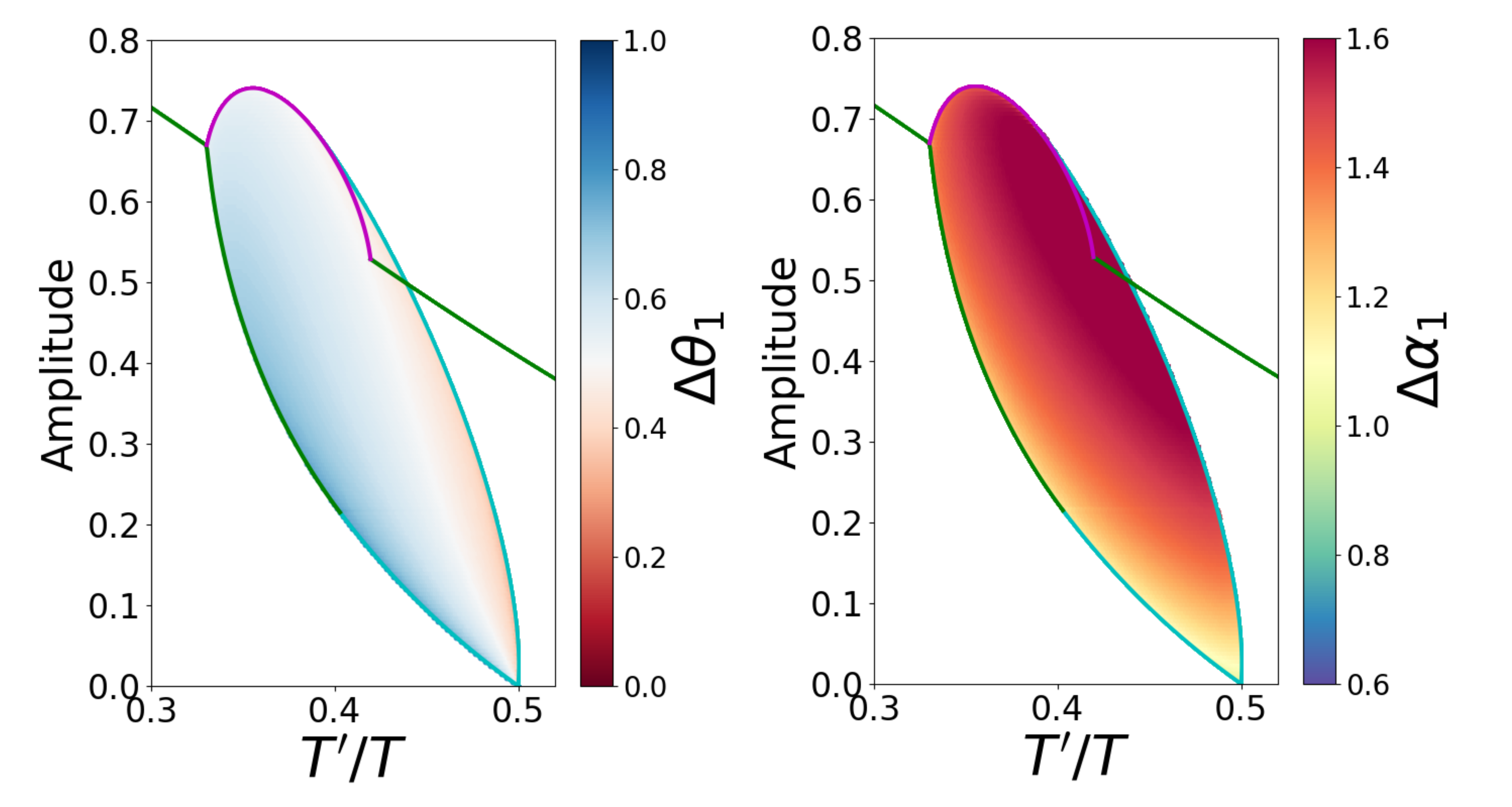} 
	\caption{For the 1:2 phase-locking region we show for the input $I_1$ the phase difference $\Delta \theta_1$ (left) and the amplitude increase factor $\Delta \alpha_1$ defined in Eq.~\eqref{eq:firstInputMagnitudes} (right).}
	%For the 1:2 phase-locking area we show: (Left panel) the normalized time difference $\Delta \theta_1$ (see Eq.~\eqref{eq:timeDifference11}) between the max of the perturbation $\theta_{pert}$ and the max of the inhibitory population $\theta_{inh}$ for the first input. (Right panel) the maximum of the oscillatory activity of the excitatory population $\alpha^A_{max}$, normalized by $\alpha^0_{max}$ the maximum of the oscillatory activity of the unperturbed excitatory population. }
	\label{fig:phases12}
\end{figure}

\begin{figure}
	\centering
	\includegraphics[width=1\linewidth]{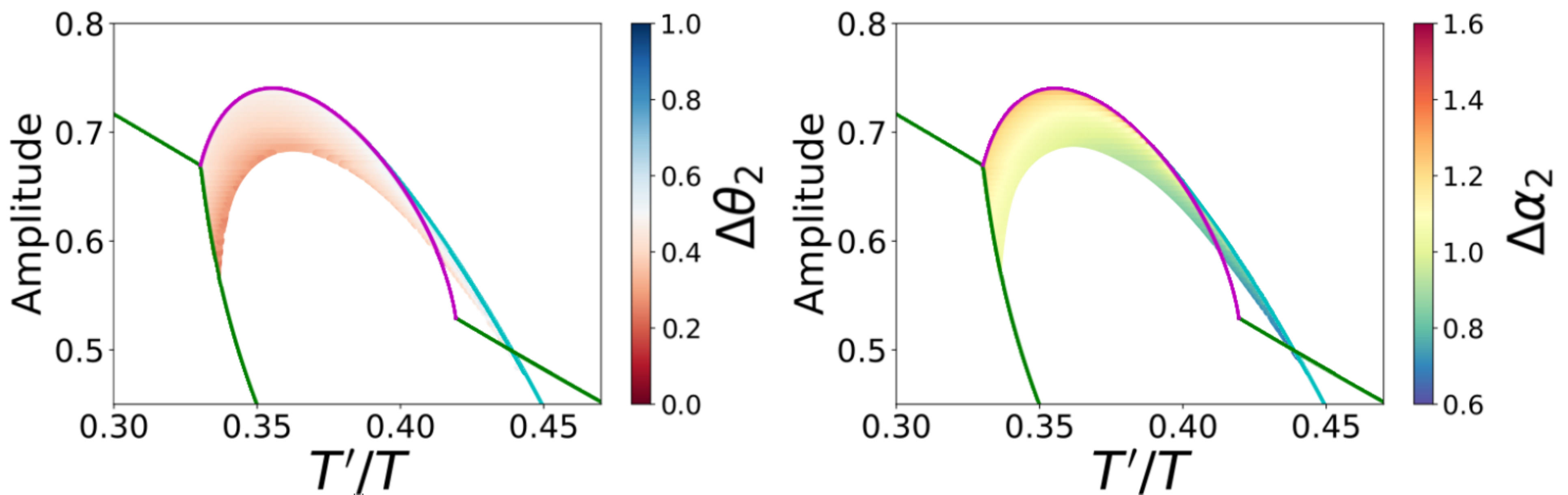} 
	\caption{For the 1:2 phase-locking region we show for the input $I_2$, whenever the excitatory/inhibitory activity shows two local maxima, the phase difference $\Delta \theta_2$ (left) and the amplitude increase factor $\Delta \alpha_2$ defined in Eq.~\eqref{eq:secondInputMagnitudes} (right).
	}
	%For the 1:2 phase locked area we show: (Left panel) the normalized time difference $\Delta \theta_2$ (see Eq.~\eqref{eq:timeDifference11}) between the max of the perturbation $\theta_{pert}$ and the max of the inhibitory population $\theta_{inh}$ for the second input. (Right panel) the maximum of the oscillatory activity of the excitatory population $\alpha^A_{max}$, normalized by $\alpha^0_{max}$ the maximum of the oscillatory activity of the unperturbed excitatory population.}
	\label{fig:amplitudes12}
\end{figure}

\begin{figure}[H]
	\centering
	\includegraphics[width=1\linewidth]{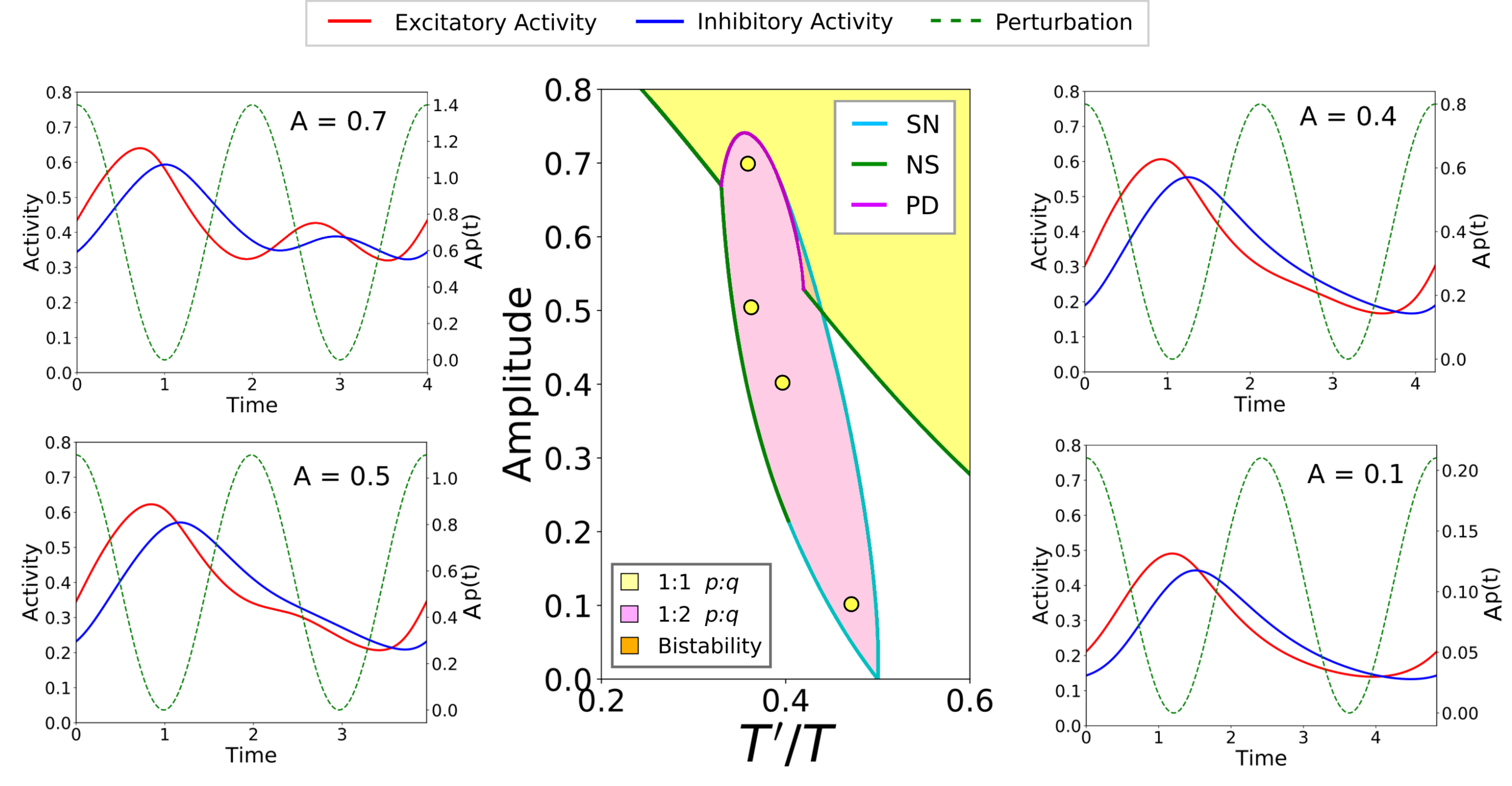} 
	\caption{For the 1:2 phase locking region we show some phase locked states as the amplitude of the perturbation is increased. From lower to higher amplitude values, the frequency values are 0.46, 0.4, 0.37 and 0.38, respectively. }
	\label{fig:phases12b}
\end{figure}

Fig~\ref{fig:phases12} shows the magnitudes $\Delta \theta_1$ and $\Delta \alpha_1$ defined in \eqref{eq:firstInputMagnitudes}, for the first input $I_1$. Observe that as $I_1$ was defined as always preceding the main inhibitory response, then $0 < \Delta \theta_1 < 1$ (see Fig.~\ref{fig:phases12} left panel), so similarly to the 1:1 phase-locking case, the effect of this input is to increase the activity of the excitatory population ($\Delta \alpha_1 > 1$) (see right panel in Fig.~\ref{fig:phases12}).  By contrast, the second input only elicits a response of the receiving population for large values of the amplitude (see coloured region in Fig.~\ref{fig:amplitudes12} and Fig~\ref{fig:phases12b}), and this is smaller than the one produced by the first input. Indeed, both inputs only elicit the same response just at the upper boundary of the 1:2 phase-locking region.

%\marginpar{En aquesta Figura 15 i en la 16 he intentat mantenir el mateix codi de colors pero no se si ajuda.}

In conclusion, the 1:2 phase-locking pattern naturally establishes a stable phase relationship so that one of the inputs enhances the response of the excitatory neurons while preventing the second one to elicit a response (except at the upper boundary of the 1:2 phase-locking region). Notice that because of the symmetry of the problem, no input is preferred, so phase-shifts can change the input selected for effective communication.

\subsection{Bistable regions}

The analysis in Section~\ref{sec:bifAnalisys} revealed the existence of different bistable regions which can be interpreted in terms of the CTC framework. Bistability suggests that, for a given input, the population may operate in different regimes depending on the initial conditions (which in fact correspond to the initial phase difference between oscillators). More interestingly, the bistability regions that we have found can generate situations in which two different synchronous regimes or co-existence of synchronous and asynchronous regimes are possible. Namely,
\begin{itemize}
	\item Bistability between a 2-periodic orbit and an invariant curve without fixed or 2-periodic points for the map $F_A$ (panel D in Fig.~\ref{fig:areaA4} and panels E and I in Fig.~\ref{fig:areaA5}).
	\item Bistability between a 2-periodic orbit and a fixed point of $F_A$ (panel E in Fig.~\ref{fig:areaA4}).
	\item Bistability between a fixed point and an invariant curve without fixed points of the map $F_A$ (panel C in Fig.~\ref{fig:areaA6}).
	\item Bistability between two stable fixed points of the map $F_A$ (panels D and E in Fig.~\ref{fig:areaA6} and panels C and D in Fig.~\ref{fig:areaA7}).
\end{itemize}

\begin{figure}[H]
	\centering
	\includegraphics[width=1\linewidth]{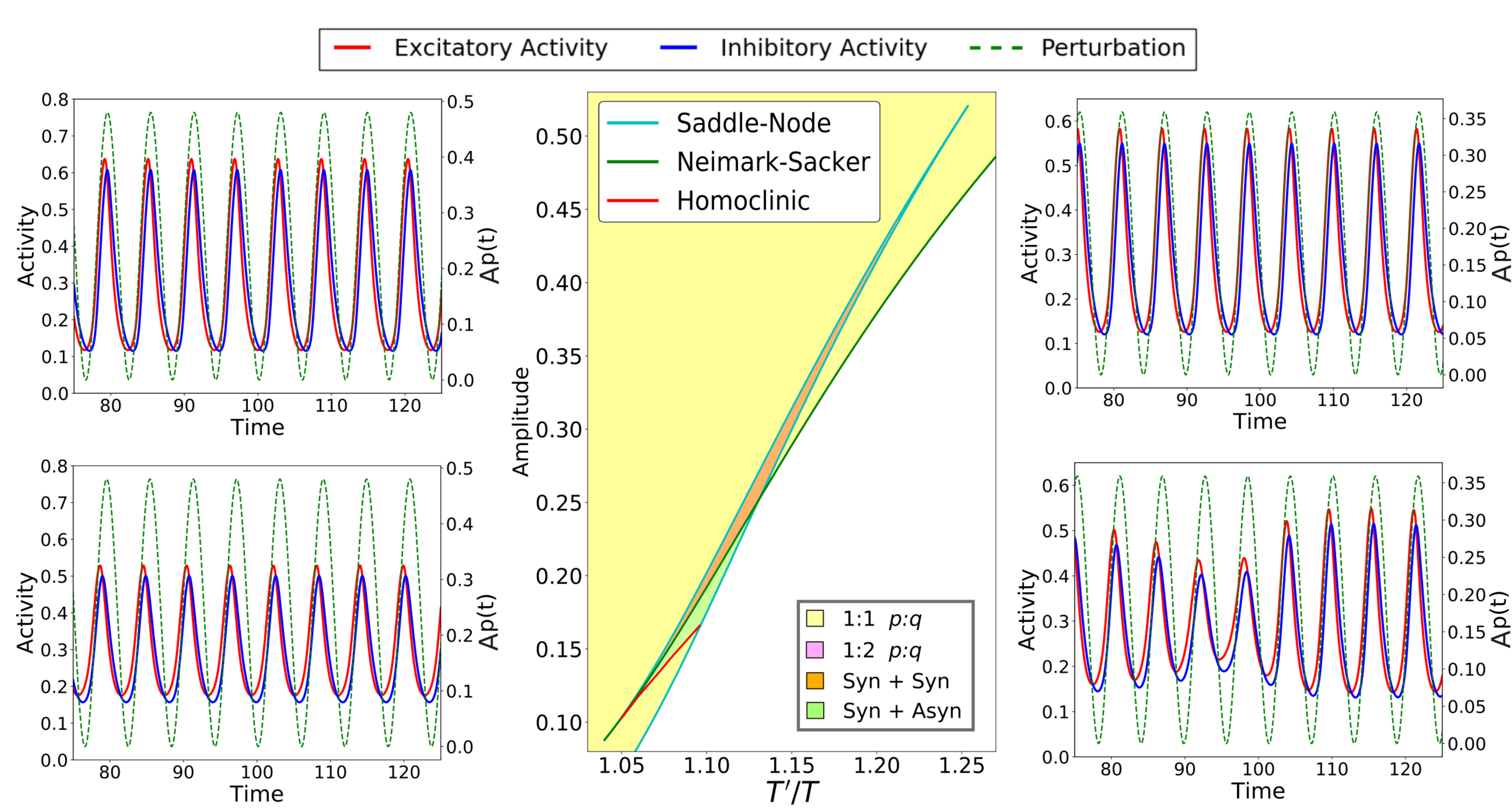} 
	\caption{For the region of bistability in the 1:1 phase-locking area we show examples of bistability between synchronous solutions (left column) and between synchronous and asynchronous solutions (right column). Central panel shows a zoom of the bifurcation diagram in Fig.~\ref{fig:HBbifDiagLarge}. The bistable dynamics in the left column (Syn + Syn) can be found in the orange region of the central panel corresponding to regions D and E in Fig.~\ref{fig:areaA6}. Alternatively, the bistable dynamics in the right column (Syn + Asyn) can be found in the green region of the central panel corresponding to region C in Fig.~\ref{fig:areaA6}.}
	\label{fig:bistability11}
\end{figure}

%
%As one can see this wide variety of bistable phenomena

\begin{figure}[t]
	\centering
	\includegraphics[width=1\linewidth]{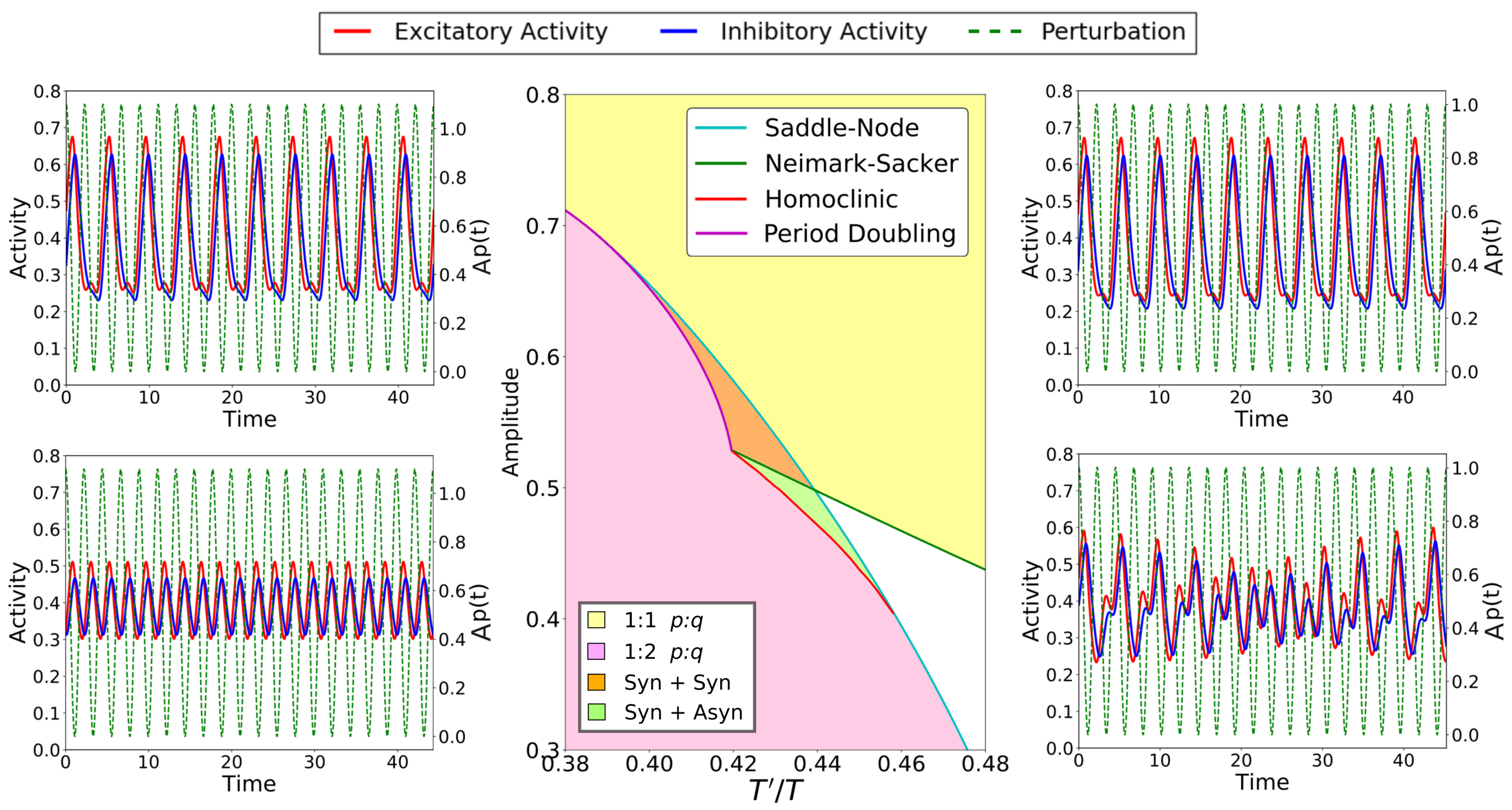} 
	\caption{For the region of bistability in the 1:2 phase-locking region we show examples of bistability between synchronous solutions (left column) and between synchronous and asynchronous solutions (right column). Central panel shows a zoom of the bifurcation diagram in Fig.~\ref{fig:HBbifDiagLarge}. The bistable dynamics in the left column (Syn + Syn) can be found in the orange region of the central panel corresponding to region E in Fig.~\ref{fig:areaA4}. Alternatively, the bistable dynamics in the right column (Syn + Asyn) can be found into the green region of the central panel corresponding to region D in Fig.~\ref{fig:areaA4}.}
	\label{fig:bistability12}
\end{figure}

Figures \ref{fig:bistability11} and \ref{fig:bistability12} show the main bistable regions. In particular, bistable situations between fixed points of the stroboscopic map imply a defined phase locking relationship, suggesting that there can exist different encodings of the input by the receiving population depending on the initial phase difference, as Fig~\ref{fig:bistability11} left illustrates. In this case, one of the solutions shows a larger variation in the activity of the E cells. We also observe bistability between 1:1 and 1:2 entrainment, see Fig.\ref{fig:bistability12} left, where the receiving population can either select only one input or respond to both at the price of reducing its effects. 

By contrast, bistable situations between fixed points and attracting invariant curves, as illustrated in Fig~\ref{fig:bistability11} right and Fig.~\ref{fig:bistability12} right, suggest that there might exist or not coherence between the emitting and receiving neural groups depending on the initial conditions. The absence of coherence (asynchronous regimes) prevents the communication between them. 

\section{Discussion}\label{sec:discussion}
%\section{Conclusions}\label{sec:lastSectionC2}

In this paper we have introduced a mathematical framework based on a phenomenological description of the population activity to study some aspects of the CTC theory, namely coherence or selective communication. Our approach considers an oscillating population of excitatory and inhibitory neurons described by the Wilson-Cowan equations, which models the receiving population, submitted to an external time-periodic input, which models the effect of the emitting population. By varying the amplitude and the frequency of the external forcing, we studied the phase-locking regions between the forcing and the system, and interpreted the dynamics in terms of the CTC theory.

To do so, we considered the stroboscopic map $F_A$ and computed the bifurcation diagram of the fixed points in terms of the amplitude and the frequency. We have focused in the regions corresponding to 1:1 and 1:2 phase locking since they are the largest ones and have stronger implications for CTC theory. In general, our analysis revealed the existence in the parameter space of only one attracting object, either a fixed point for the map $F_A$ (or $F^2_A$) or an invariant curve without fixed points on it, which correspond to stable synchronous or asynchronous regimes, respectively. By performing a detailed analysis of the boundaries of these regions, we have found rich dynamics, like bistability between invariant objects.

Once we have identified the stable phase-locked states of the system, we have analysed those aspects of the dynamics that have important implications for the CTC theory. In particular, the phase relationship between the inhibition and the input and 
the increase in the activity of the target excitatory population due to the external input. Indeed, for an effective communication (a positive effect of the input onto the activity of the excitatory target population), the theory requires the input to precede activation of the inhibition, since the inhibition may partially or totally silence the input effect. In general, we have found that the entrainment of the postsynaptic population to the rhythmic input from the presynaptic population naturally sets up a phase relationship that is optimal for CTC, in the sense that the input precedes the inhibition, leading to an effective communication. Interestingly, we have found that near the borders of the phase-locking region, where the transition from synchronous to asynchronous dynamics occurs, the general tendency is reversed and the inhibition precedes the input. This result suggests a relationship between the loss of effective communication and the loss of phase-locking.

By repeating the analysis for the 1:2 phase-locking region, we explored a different aspect of the CTC theory, which is selective communication, that is, for a population that receives two identical inputs from different sources, how it can respond to one while ignoring the other \cite{fries2015rhythms}. To do so, for this region, we interpreted the input to the receiving population as the sum of two identical inputs arriving in anti-phase (see Fig~\ref{fig:ctcFramework_v2}), and observed that the phase automatically sets up so that one of the inputs precedes and the other one follows the inhibitory action. Our results confirm the hypothesis that the input following the inhibitory action has almost no effect onto the target population. 

Moreover, we have found regions with bistability. Bistability suggests that depending on the initial conditions, the population may operate in different regimes without changing the structure or the connections of the network. 
Thus, bistability between synchronous and asynchronous solutions suggests that communication between brain areas cannot be predicted by the actual network but depends on the current state of the network (initial conditions), indicating that communication between neuronal populations can be switched on and off by means of possibly top-down influences \cite{fries2015rhythms}.

%	\subsection{Future Perspectives}	

	 	%In this paper we have shown how a simplified setting, consisting on the periodic forcing of a population firing rate model, can extract the essence of CTC theory. 
 	%As the only requirement of the mathematical analysis is the periodicity of the input, it opens the door to explore other hypothesis and propositions of the CTC theory. 
 	%We remark that our approach does not focus on a particular aspect but on the patterns that emerge from the particular mathematical framework 
	%that we propose for CTC, so we have identified selective communication, proper phase-locking and bistability.
	%(in fact we refer the reader to the Appendix \ref{ap:nonSin} for a discussion about weather this result might or not depend on the sinusoidal type of input chosen)
	Thus, we stress that thanks to the low-dimensionality of our system, we have been able to obtain a bifurcation diagram which provides a broad picture of the dynamics in a large parameter space. Thus, our results confirm previous computational results on spiking networks, while they provide new results that suggest further research. More precisely, our study corroborates the general result for network approaches \cite{fries2015rhythms, cannon2014neurosystems}, in which the phase established between the input and the receiving network is optimal for CTC. Moreover, we have also observed that selective communication occurs and depends on the amplitude and frequency (see also \cite{GielenKZ10, borgers2008gamma}). Moreover, because of the wide range of parameters explored, we have found new interesting results. Namely, we have been able to detect the break up of the phase-locking patterns and relate them to the phase relationship between excitation and inhibition, as well as to detect the possibility of having bistability between synchronous regimes or synchronous and asynchronous regimes, which can motivate future research.
	
	The CTC theory proposes several hypothesis and we have not included all of them here.  Other papers in the neuroscience literature regarding the mathematical implementation of the CTC theory, focus on different aspects than the ones considered here. As the only requirement of our mathematical analysis is the periodicity of the input, it can be applied to explore other hypothesis and propositions of the CTC theory. For instance, some studies related to our problem have explored the coherence (in the sense of width) of the input in selective communication \cite{borgers2008gamma, GielenKZ10}. Instead of a sinusoidal input, we could have considered other types of functions for the input, like Gaussian-shaped with a parameter that controls the width of the input coherence (see, for instance, \cite{borgers2008gamma, cannon2014neurosystems}) to study this feature of CTC by means of our setting. We refer the reader to the Appendix \ref{ap:nonSin} for preliminary results on other types of inputs. Moreover, we could have also explored the implications of inputs to
	the inhibitory cells (see, for instance, \cite{veltz2015}), as they have been also proposed to be a mechanisms of phase-shifting in models of cortical networks \cite{tiesinga2010}. Indeed, regarding the role of inhibition in the CTC, the Wilson-Cowan equations that we used describe oscillations across the PING (Pyramidal Interneuron Network Gamma) mechanism \cite{devalle2017firing}. As the ING (Interneuron Network Gamma) mechanism can also account for gamma E-I oscillations \cite{tiesinga2000robust, wang1996gamma}, we can use our setting to explore the implications of ING mechanism in the CTC \cite{tiesinga2009cortical}. Furthermore the extension of our setting to networks of two populations, can lead to the exploration of bi-directional communication \cite{witt2013controlling, battaglia2012dynamic,dumont2019macroscopic, perez2019uncoupling} instead of the unidirectional communication that we have explored in this paper. 
	
	Finally, the model can be adjusted in order to provide a more biologically-based mathematical framework. As we stated in the introduction, the framework that we introduce in this paper is an alternative to large-dimensional network approaches to the CTC.  A macroscopic observable measuring the mean rate of such neuronal networks has been described by means of PDE equations such as the Fokker-Planck equation \cite{BrunelH99} or the refractory density equation \cite{gerstner2002spiking} which have been used to describe emerging oscillations in such neuronal networks.  Interestingly, recently several groups have developed exact firing rate models solving the Fokker-Planck PDE equation, thus leading to new neural mass models described by ODEs \cite{ montbrio2015macroscopic, Coombes2019}, to which we can apply the methodology described in this paper. As the results from these models can be automatically checked on the spiking network, these new models appear as a feasible and more realistic alternative to heuristic equations such as the Wilson-Cowan equations that we use. Furthermore, independently of the rate model which is being used, the addition of synaptic equations can account for the description of realistic gamma band scenarios in which our scheme can also be applied \cite{Keeley19}. 
	
	Besides the neuroscience considerations, we would like also to emphasize several mathematical implications of our work. Perturbations of non-linear oscillators have been extensively studied in the mathematical literature \cite{AshwinCN16, hoppensteadt2012weakly}. In particular, this paper explores the dynamics of a perturbed oscillator close to a Hopf bifurcation. 
	Our results are based on a rigorous numerical study of a particular system (the Wilson-Cowan equations), which does not have any restrictions regarding the size of the amplitude or the forcing frequency \cite{ZhangG11}. The results obtained numerically match the theoretical predictions for perturbations of a Hopf bifurcation obtained in \cite{gambaudo1985} using the normal form. 
	We highlight here that our method does not require the system to be close to the Hopf bifurcation. Indeed, we also
	have applied it to the case close to the SNIC bifurcation (see Appendix~\ref{ap:bd_snic}). Interestingly, although the regions of strong resonances show a different shape, we obtain the same bifurcations of fixed points.
	A detailed study of the implications for CTC theory of this scenario is an interesting topic for future research.

	In conclusion, we have shown that a simplified setting that extracts the essence of the CTC theory allows a basic understanding of the processes involved in the generation of communication through rhythms according to the CTC theory. We expect that it would shed light into the field and open the door for future studies.

\appendix

\section{Numerical computation of the bifurcation Diagram}\label{sec:bifAnalysis}

In this Section we provide a brief description of the numerical procedure used to compute the bifurcations of the fixed points of the map $F_A$ defined in~\eqref{eq:genericMap}. 
We highlight that we have developed our own numerical software in Python to compute the bifurcation diagram instead of relying on existing software packages, thus providing more control on the 
calculations performed. We consider three types of bifurcations of fixed points: Saddle-Node (\textit{SN}), Period Doubling (\textit{PD}) and Neimark-Sacker (\textit{NS}) (see \cite{kuznetsov2013elements}).

We first note that the parameters that we will vary for the map $F_A$ in \eqref{eq:genericMap} will be the amplitude $A$ and the period $T'$ of the perturbation. 
The computation of the bifurcation curves is based on the numerical methods provided in \cite{simo1990}. Here we summarize the main steps of the procedure.

We first assume that $A$ is fixed and we look for a point $x \in \mathbb{R}^2$ and a period $T' \in \mathbb{R}$ that satisfy that $x$ is a fixed point of the map $F_A$ together with a bifurcation condition $\Phi_{BIF}(x)=0$.

To set the mathematical formalism, we consider the extended system consisting of system~\eqref{eq:WCsys} (that we will denote generically $\dot{x}=X(x,t,T')$ where $x=(x_1,x_2) \in \mathbb{R}^ 2$) 
with the extra equation $\dot{T}'=0$. Let us denote $\tilde \phi_A (t;t_0,x,T')$ the flow of the extended system and let us introduce the map 
%\textcolor{blue}{To set the mathematical formalism, we need to consider the extended system~\eqref{eq:WCsys} (see Remark \ref{rm:remarkGemma}). The extended system will be denoted generically $\dot{x}=X(x,t,T')$ where $x=(x_1,x_2) \in \mathbb{R}^ 2$, and it consists on the original system~\eqref{eq:WCsys}  
%and the extra equation $\dot{T}'=0$. If we denote by $\tilde \phi_A (t;t_0,x,T')$ the flow of the extended system we can define its corresponding stroboscopic map}

\begin{eqnarray}\label{eq:genericMapExt}
\tilde{F}_{A}:&\mathbb{R}^{3} &\to \mathbb{R}^{3}, \notag \\
&z=(x,T') &\to \tilde{F}_A(z) = (\tilde{F}^x_A(z),\tilde{F}^{T'}_A(z))=\tilde{\phi}_A(T'; z),
\end{eqnarray}
Notice that we have set $t_0=0$ and abusing of notation we have removed the dependence on $t_0$ from the expression of the flow $\tilde{\phi}_A$. 
From now on we will also remove the subscript $A$ to avoid stodgy notation.
The superscript $\tilde F^w$ refers to the $w$-component of the map $\tilde F$, where $w=x_1,x_2,T'$.

\begin{remark}\label{rm:remarkGemma}
We consider the extended system with the trivial equation $\dot{T'}=0$ because, as we will see later, we need to know how the 
solutions of system~\eqref{eq:WCsys} vary with respect to the parameter $T'$.
\end{remark}

Thus, to detect the bifurcation values we need to look for zeroes of the equation
\begin{equation}\label{eq:bifSystem}
G(x, T') =
\begin{cases}
\tilde F^x(x, T') - x = 0 \\
\Phi_{BIF}(x, T') = 0.
\end{cases}
\end{equation}

The conditions which must be satisfied at the bifurcation values for a Saddle-Node (\textit{SN}), Period Doubling (\textit{PD}) and Neimark-Sacker (\textit{NS}) bifurcations are, respectively,
\begin{equation}\label{eq:lesCondicions}
\begin{aligned}
	\Phi_{SN}(x, T') & = \textrm{det}(D_x \tilde F^{x}(x, T') - Id_2) = 0, \\
	\Phi_{PD}(x, T') & = \textrm{det}(D_x \tilde F^{x}(x, T') + Id_2) = 0, \\
	\Phi_{NS}(x, T') & = \textrm{det}(D_x \tilde F^{x}(x, T')) - 1 = 0,
\end{aligned}
\end{equation}
where we denote by $D_x \tilde F^{x}$ the Jacobian matrix of the map $\tilde F^x$  restricted to the first two components. 

We have implemented a Newton method to solve system~\eqref{eq:bifSystem}. Suppose we have an approximate solution $z = (x, T')$ of system \eqref{eq:bifSystem} and we look for an improved solution $z^* = z + \Delta z$. Straightforward calculations show that $\Delta z$ has to satisfy:
\begin{equation}\label{eq:newtonCond}
\overbracket{
% \begin{bmatrix}
% \dfrac{\partial F^{x_1}}{\partial x_1}(z)-1 & \dfrac{\partial F^{x_1}}{\partial x_2}(z) & \dfrac{\partial F^{x_1}}{\partial T'}(z) \\
% \dfrac{\partial F^{x_2}}{\partial x_1}(z) & \dfrac{\partial F^{x_2}}{\partial x_2}(z)-1 & \dfrac{\partial F^{x_2}}{\partial T'}(z) \\
% \dfrac{\partial \Phi_{BIF}}{\partial x_1}(z) & \dfrac{\partial \Phi_{BIF}}{\partial x_2}(z) & \dfrac{\partial \Phi_{BIF}}{\partial T'}(z)
% \end{bmatrix}
\begin{bmatrix}
\partial_{x_1} \tilde F^{x_1}(z)-1 & \partial_{x_2} \tilde F^{x_1}(z) & \partial_{T'} \tilde F^{x_1}(z) \\
\partial_{x_1} \tilde F^{x_2}(z) & \partial_{x_2} \tilde F^{x_2}(z)-1 & \partial_{T'} \tilde F^{x_2}(z) \\
\partial_{x_1} \Phi_{BIF}(z) & \partial_{x_2} \Phi_{BIF}(z) & \partial_{T'} \Phi_{BIF}(z)
\end{bmatrix}
}^{DG(z)}
\overbracket{
\begin{bmatrix}
\Delta x_1 \\
\Delta x_2 \\
\Delta T'
\end{bmatrix}}^{\Delta z} = - 
\overbracket{
\begin{bmatrix}
\tilde F^{x_1}(z) - x_1 \\
\tilde F^{x_2}(z) - x_2 \\
\Phi_{BIF}(z)
\end{bmatrix}}^{E}
\end{equation}
Notice that for the first two rows of the matrix $DG$ we have
\begin{equation}
\begin{aligned}
\partial_{x_j} \tilde F^{x_k}(z) &= \partial_{x_j} \tilde{\phi}^{x_k}(T'; z),\\
\partial_{T'} \tilde F^{x_k}(z) &= \partial_{T'} (\tilde{\phi}^{x_k}(T'; z)) = \frac{d \tilde{\phi}^{x_k}}{dt} (t; z)|_{t=T'} + \frac{\partial \tilde{\phi}^{x_k}}{\partial{T'}}(t; x,T')|_{t=T'}\\
&=X^{x_k}(\tilde{\phi}(T'; z))+\frac{\partial \tilde{\phi}^{x_k}}{\partial{T'}}(t; x,T')|_{t=T'} \\
\end{aligned}
\end{equation}
for $j,k = 1, 2$. In order to obtain the derivatives of the flow $\tilde{\phi}$ with respect to initial conditions $(x_1,x_2,T')$ at time $T'$, we need to integrate
the first variational equations given by
\begin{equation}
\frac{d}{dt} D_z \tilde \phi (t;z) = A(t) D_z \tilde \phi (t;z) 
\end{equation}
with initial condition $D_z \tilde \phi(0;z) = Id_3$,
where
\[A(t)=\begin{bmatrix}
\partial_{x_1} X^{x_1} & \partial_{x_2} X^{x_1} & \partial_{T'} X^{x_1} \\
\partial_{x_1} X^{x_2} & \partial_{x_2} X^{x_2} & \partial_{T'} X^{x_2} \\
0 & 0 & 0 
\end{bmatrix}
_{|\tilde \phi (t,z)}
\]
Recall that these equations need to be integrated together with the flow $\tilde \phi (t,z)$ up to time $T'$.
% \begin{bmatrix}
% \dot{\tilde \phi}^{x_1}_{x_1} & \dot{\phi}^{x_1}_{x_2} & \dot{\phi}^{x_1}_{T'} \\
% \dot{\tilde \phi}^{x_2}_{x_1} & \dot{\phi}^{x_2}_{x_2} & \dot{\phi}^{x_2}_{T'} \\
% \dot{\tilde \phi}^{T'}_{x_1} &
% \dot{\phi}^{T'}_{x_2} & \dot{\phi}^{T'}_{T'}
% \end{bmatrix} =  
% \begin{bmatrix}
% D_{x_1} X^{x_1}() & D_{x_2} X^{x_1}(z) & D_{T'} X^{x_1}(z) \\
% D_{x_1} X^{x_2}(z) & D_{x_2} X^{x_2}(z) & D_{T'} X^{x_2}(z) \\
% D_{x_1} X^{T'}(z) & D_{x_2} X^{T'}(z) & D_{T'} X^{T'}(z) 
% \end{bmatrix}
% \begin{bmatrix}
% \phi^{x_1}_{x_1} & \phi^{x_1}_{x_2} & \phi^{x_1}_{T'} \\
% \phi^{x_2}_{x_1} & \phi^{x_2}_{x_2} & \phi^{x_2}_{T'} \\
% \phi^{T'}_{x_1} &
% \phi^{T'}_{x_2} & \phi^{T'}_{T'}
% \end{bmatrix} 
% \end{equation}
%

Finally, one needs to compute the terms in the third row of $DG$ in \eqref{eq:newtonCond}. The exact expression depends on the bifurcation for which we look for (see Eq. \eqref{eq:lesCondicions}).
Next, we will derive the expression for the $SN$ case to illustrate the method, and the other cases are analogous. The determinant of $D_x \tilde F^x - Id_2$ writes as
\begin{equation}
\Phi_{SN} = \det(D_x \tilde F^x - Id_2) = (\tilde \phi^{x_1}_{x_1} - 1)(\tilde \phi^{x_2}_{x_2} - 1) - \tilde\phi^{x_2}_{x_1}\tilde\phi^{x_1}_{x_2},
\end{equation}
where $\tilde \phi^{x_k}_{x_j}=\partial_{x_j} \tilde \phi^{x_k}(T';x,T')$ and therefore
\begin{equation}\label{eq:secondOrder}\begin{aligned}
\partial_{x_1} \Phi_{SN} = & \tilde\phi^{x_1}_{x_1x_1}\tilde\phi^{x_2}_{x_2} + \tilde\phi^{x_1}_{x_1}\tilde\phi^{x_2}_{x_2x_1} - \tilde\phi^{x_1}_{x_1x_1} - \tilde\phi^{x_2}_{x_2x_1} - \tilde\phi^{x_2}_{x_1x_1}\tilde\phi^{x_1}_{x_2} - \tilde\phi^{x_2}_{x_1}\tilde\phi^{x_1}_{x_2x_1}\\
\partial_{x_2} \Phi_{SN} = & \tilde\phi^{x_1}_{x_1x_2}\tilde\phi^{x_2}_{x_2} + \tilde\phi^{x_1}_{x_1}\tilde\phi^{x_2}_{x_2x_2} - \tilde\phi^{x_1}_{x_1x_2} - \tilde\phi^{x_2}_{x_2x_2} - \tilde\phi^{x_2}_{x_1x_2}\tilde\phi^{x_1}_{x_2} - \tilde\phi^{x_2}_{x_1}\tilde\phi^{x_1}_{x_2x_2}\\
\partial_{T'} \Phi_{SN} = & \tilde\phi^{x_1}_{x_1T'}\tilde\phi^{x_2}_{x_2} + \tilde\phi^{x_1}_{x_1}\tilde\phi^{x_2}_{x_2T'} - \tilde\phi^{x_1}_{x_1T'} - \tilde\phi^{x_2}_{x_2T'} - \tilde\phi^{x_2}_{x_1T'}\tilde\phi^{x_1}_{x_2} - \tilde\phi^{x_2}_{x_1}\tilde\phi^{x_1}_{x_2T'}
\end{aligned}
\end{equation}
where $\tilde\phi^{z_k}_{z_i z_j} = \frac{\partial^2} {\partial_{z_j} \partial_{z_i}} \tilde \phi^{z_k}(T';z)$ for $k=1,2$ and $i,j=1,2,3$. Notice that $z_3 = T'$.

The computation of the terms in equations~\eqref{eq:secondOrder} requires to solve the second order variational equations, which are given in components as
\begin{equation}
\frac{d}{dt} \tilde \phi^{z_k}_{z_i z_j}(t;z) = \sum^3_{p,q = 1} \partial_{z_p z_q} X^{z_k}(\tilde\phi(t;z))\frac{\partial \tilde\phi^{z_p}(t;z)}{\partial z_i} \frac{\partial \tilde\phi^{z_p}(t;z)}{\partial z_j} + \sum^3_{p=1} \partial_{z_p} X^{z_k}(\tilde\phi(t;z))\frac{\partial^2 \tilde\phi^{z_p}(t;z)}{\partial z_i \partial z_j},
\end{equation}
with initial condition $\tilde \phi^{z_k}_{z_i z_j}(0;z)=0$ and $k=1,2$, $i,j=1,2,3$.
%

%Once one has a point $z^*$ good enough, the procedure to continue to bifurcation curve consists in select a new value for the amplitude $A_{new} = A + \Delta_A$, give the new point $z^*$ as initial seed and repeat the above procedure.\\
 
Once a point $z^*$ satisfies \eqref{eq:bifSystem} with the established tolerance (in our case $10^{-8}$), the procedure to continue the bifurcation curve consists in selecting a new value for the amplitude $A_{new} = A + \Delta_A$,
use the computed point $z^*$ as initial seed, and repeat the above procedure for the new value $A_{new}$. This procedure --which is the one used in this manuscript-- requires to change the sign of $\Delta_A$ by hand when the derivative of the bifurcation curve with respect to $A$ is zero. Nevertheless, alternative strategies as the Lagrange multipliers or the pseudo arclength method (see \cite{simo1990, allgower2003introduction}, respectively) can be used so the tangent vector $v$ to the bifurcation curve evaluated at $z^*$ is obtained and can be used to provide $z^* + \epsilon v$ as initial seed for $\epsilon$ small enough. \\

\textbf{Homoclinic bifurcation}\\

In this Section, we explain the method  used to compute the homoclinic bifurcation curve in Section \ref{sec:bifAnalisys}. The crossing of an homoclinic bifurcation implies the appearance/disappearance of an attracting invariant curve which surrounds the unstable focus $P_1$. Therefore, the crossing of this bifurcation implies a qualitative change in the asymptotic solutions of system \eqref{eq:genericMap} when using an initial condition near the unstable focus $P_1$. More precisely, in the case corresponding to ``Dynamics on the right hand side of the 1:2 phase-locking region'' (see Fig.~\ref{fig:areaA4}), if an attracting invariant curve exists, an initial condition near the unstable focus $P_1$ will tend to the invariant curve for a large enough time of integration (see panel D in Fig.~\ref{fig:areaA4}). Otherwise, the 2-periodic orbit $(P_2, P_4)$ will be the asymptotic solution (see panel B' in Fig.~\ref{fig:areaA4}). Similarly, in the case ``Dynamics on the bottom right of the 1:1 phase-locking region'' (see Fig.~\ref{fig:areaA6}), if an attracting invariant curve exists, an initial condition near the unstable focus $P_1$ will tend to the invariant curve for a large enough time of integration (see panel C in Fig.~\ref{fig:areaA6}). Otherwise, the fixed point $P_2$ will be the asymptotic solution (see panel B in Fig.~\ref{fig:areaA6}). In conclusion, in both cases, the homoclinic bifurcation curve is delimited by considering different values of the fraction $\frac{T'}{T}$, and slightly varying the amplitude while checking whether there is a qualitative change in the asymptotic solution of points near $P_1$.

\section{Bifurcation Diagram for oscillations close to a SNIC bifurcation}\label{ap:bd_snic}

\begin{figure}[H]
	\begin{minipage}[c]{0.7\textwidth}
	\includegraphics[width=1\linewidth]{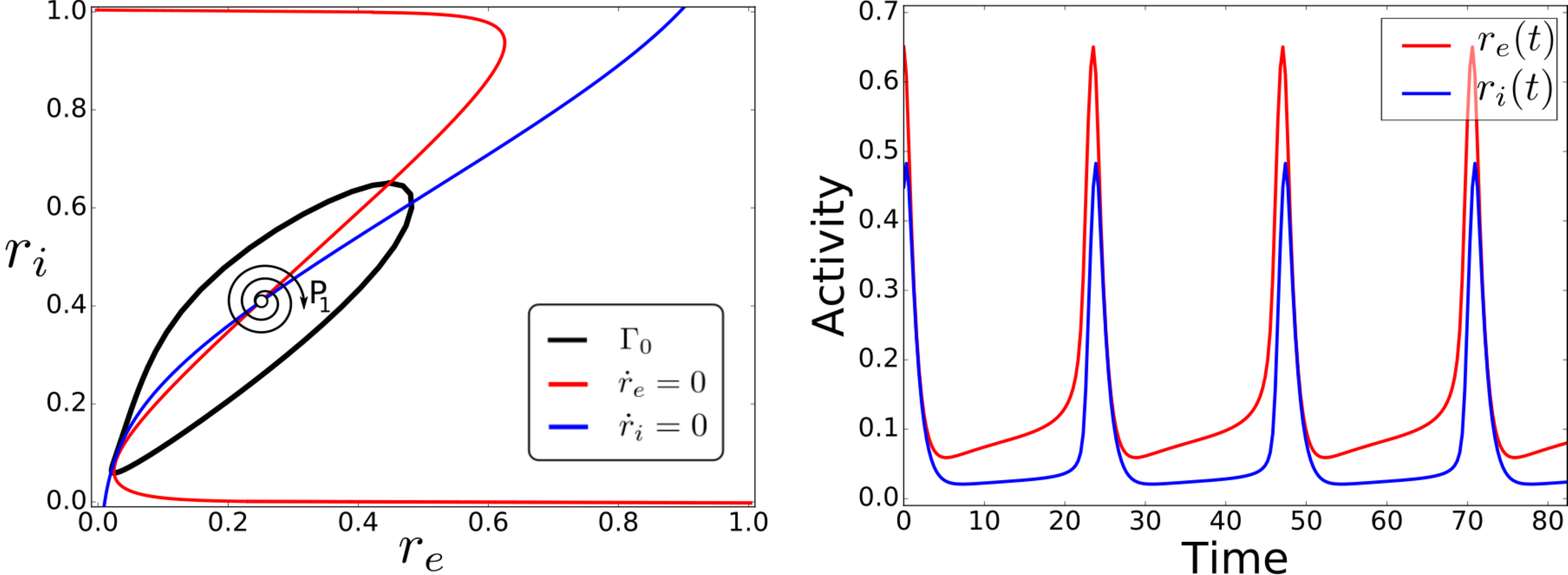} \\
	\includegraphics[width=1\linewidth]{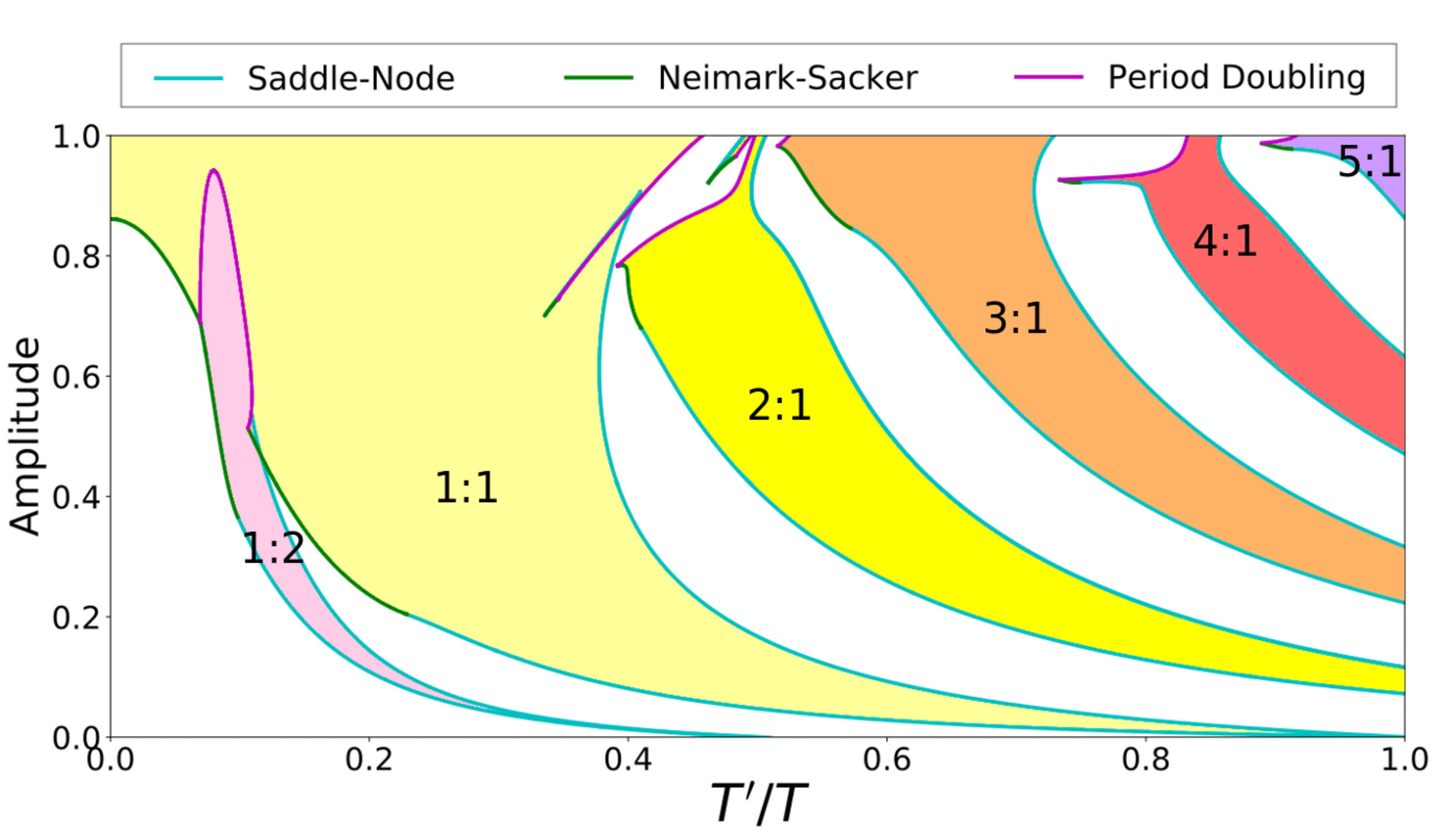}
	\end{minipage}\hfill
	\begin{minipage}[c]{0.25\textwidth}
		\caption{For the unperturbed ($A=0$) Wilson-Cowan equations \eqref{eq:WCsys} and the set of parameters $\mathcal{P}$ given in \eqref{eq:parametersChoice} we show:  
		Top-Left: Nullclines and phase space for ($P, Q$) = (1.4, -0.75). The phase space shows a limit cycle $\Gamma_0$ and an unstable focus $P_1$. 
		Top-Right: dynamics over the limit cycle $\Gamma_0$. Bottom: Bifurcation diagram for the fixed points of the stroboscopic map \eqref{eq:genericMap} of system \eqref{eq:WCsys} as the frequency and the amplitude of the perturbation 
		are varied. The coloured regions correspond to different \textit{p:q} phase locking regimes}. \label{fig:snicBifDiagram}
	\end{minipage}
\end{figure}

The unperturbed Wilson-Cowan equations \eqref{eq:WCsys} with the set of parameters $\mathcal{P}$ given in \eqref{eq:parametersChoice}, can also have oscillations which are born from a Saddle-Node on Invariant Curve (SNIC) bifurcation (see Fig.~\ref{fig:2d_bifDiag}). Thus, when we pick the values $(P,Q) = (1.4, -0.75)$, identically as to the Hopf case, the phase space for system \eqref{eq:WCsys} shows a limit cycle and an unstable focus (see Fig.~\ref{fig:snicBifDiagram}). We have carried out a numerical exploration of the bifurcations that
occur when we perturb this limit cycle close to a SNIC bifurcation. Preliminary results are shown in Figure~\ref{fig:snicBifDiagram} bottom. The regions
with \textit{p:q} phase-locking, start out of the point $(p/q ,0)$ but, as the amplitude increases, they tilt towards lower values
of $p/q$ as predicted by the Phase Response Curves \cite{ErmentroutTerman2010}. 
Indeed, the PRC for limit cycles close to a SNIC are mainly positive indicating that the
phase can only be advanced by an external excitatory perturbation.
Therefore, they can only synchronize with external inputs with higher
frequency.

Moreover, when we compare with the Hopf diagram, in both cases, the
phase-locking regions are bounded by saddle-node bifurcation curves (for
small values of $A$) and Neimark-Sacker and period-doubling bifurcation
curves for other values of the amplitude. So, our preliminary study
suggests a qualitatively similar dynamics as in the Hopf case. A thorough study is left for future work.

\section{Non-sinusoidal inputs}\label{ap:nonSin}

Our analysis has been carried out for sinusoidal inputs while the oscillations generated by the E-I network are not. Thus, in order to consider a more realistic approach, in this Section, we present preliminary results on the dynamical effects of unidirectional coupling between two Wilson-Cowan populations. Thus, 
we use the value of $r_e$ of one population as the input for the second one, which is periodic but non-sinusoidal. Mathematically, we consider the following system of equations:
\begin{equation}\label{eq:twoWCs}
\begin{aligned}
\dot{r}_{e_1} &= -r_{e_1} + S_e(c_1 r_{e_1} - c_2 r_{i_1} + P + r_{e_2}),\\
\dot{r}_{i_1} &= -r_{i_1} + S_i(c_3 r_{e_1} - c_4 r_{i_1} + Q),\\
\dot{r}_{e_2} &= -r_{e_2} + S_e(c_1 r_{e_2} - c_2 r_{i_2} + P),\\
\dot{r}_{i_2} &= -r_{i_2} + S_i(c_3 r_{e_2} - c_4 r_{i_2} + Q),
\end{aligned}
\end{equation}
where the parameters are the same for both populations and correspond to the set $\mathcal{P}$ in \eqref{eq:parametersChoice}.

\begin{figure}[H]
	\begin{minipage}[c]{0.6\textwidth}
		\includegraphics[width=1\linewidth]{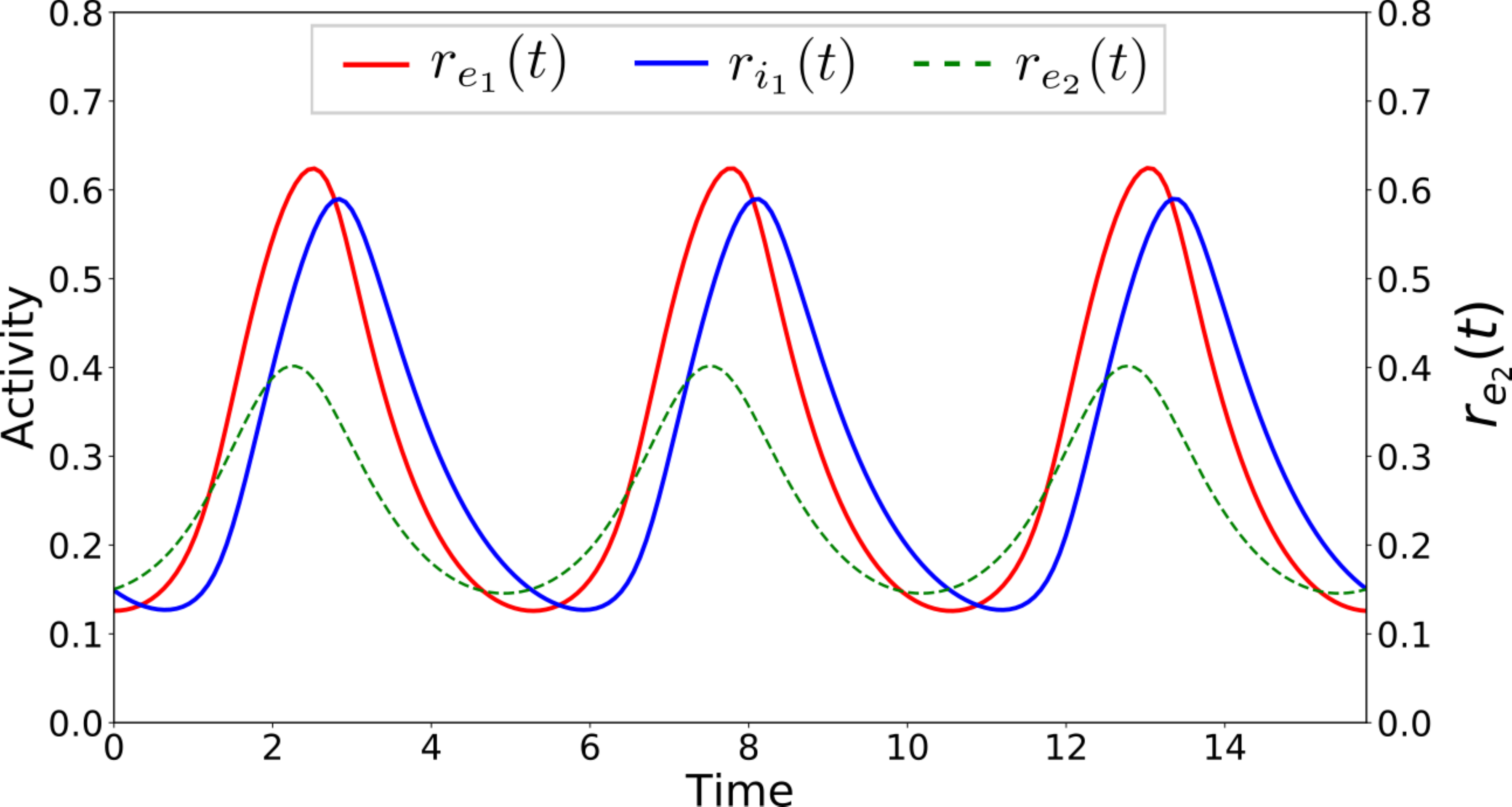}
	\end{minipage}\hfill
	\begin{minipage}[c]{0.35\textwidth}
		\caption{For the unidirectionally coupled Wilson-Cowan equations \eqref{eq:twoWCs} and the set of parameters $\mathcal{P}$ given in \eqref{eq:parametersChoice} we show 
		the firing rate dynamics of the receiving population $r_{e_1}$ and $r_{i_1}$, and the non-sinusoidal input corresponding to $r_{e_2}$. Results show that an optimal phase relationship $\Delta \theta > 0$ leading to an effective communication $\Delta \alpha > 1$ is established.} \label{fig:movRev}
	\end{minipage}
\end{figure}

When we simulate this system, the network establishes a 1:1 phase-locking and a phase relationship $\Delta \theta > 0$ that increases the amplitude of oscillations $\Delta \alpha > 1$ (see Fig.~\ref{fig:movRev}), 
in agreement with the results of the paper presented in Section~\ref{sec:sec41}. 
%and with computational studies (see \cite{cannon2014neurosystems} for instance). 
A complete investigation varying the amplitude and frequency of the input, or even the coherence (in the sense of sharpness) of the input (see \cite{cannon2014neurosystems}) is left for future work. Indeed, regarding the coherence of the input,
preliminary results with inputs of the form $p(t)=\cos^n(2 \pi t/T')$ with $n=4,8$, which have higher coherence than the input \eqref{eq:perturbationCosinus} considered in this paper, show qualitatively the same bifurcation diagram as in Fig~\ref{fig:HBbifDiagLarge} (results not shown).

%This remarks that some conclusions of this paper probably will hold for non sinusoidal inputs. Finally, as future work, we can explore bidirectional coupling by studying two identical coupled populations of Wilson-Cowan equations \cite{Borisyuk1995} (see \cite{perez2019uncoupling} for the special case near a Hopf bifurcation). 

% \begin{figure}[H]
% 	\begin{minipage}[c]{0.7\textwidth}
% 	\includegraphics[width=1\linewidth]{./images/diagramSN_v2.png} 
% 	\end{minipage}\hfill
% 	\begin{minipage}[c]{0.25\textwidth}
% 		\caption{Bifurcation diagram for the fixed points of the stroboscopic map \eqref{eq:genericMap} of system \eqref{eq:WCsys} as the frequency and the amplitude of the perturbation are varied. Solid lines correspond to bifurcations of stable fixed points.} \label{fig:snicBifDiagram}
% 	\end{minipage}
% \end{figure}

\subsection*{Acknowledgements}

This work has been partially funded by the Spanish grants MTM2015-65715-P, MDM-2014-0445, PGC2018-098676-B-100, the Catalan grant 2017SGR1049 (GH, AP, TS), the MINECO-FEDER-UE MTM-2015-71509-C2-2-R (GH), 
and the Russian Scientific Foundation Grant 14-41-00044 (TS).
GH acknowledges the RyC project RYC-2014-15866. TS is supported by the Catalan Institution for research and advanced studies via an ICREA academia price 2018.
AP acknowledges the FPI Grant from project MINECO-FEDER-UE MTM2012-31714.

\subsection*{Competing Interests}

The authors declare they have no competing interests.

\bibliographystyle{unsrt}
%NOLINEAL16 you have to edit the file end.tex, changing bibjimenez to your bibfile name should read as:
\bibliography{bibForcing} %it should be kept in your author folder

\end{document}